\documentclass[12pt,preprint]{emulateapj}

\usepackage{natbib}
\usepackage{graphics,graphicx}
\usepackage{amsmath}
\usepackage{lineno}

\newcommand{\g}{$\gamma$}

\newcommand{\Cha}{{\it Chandra~}}

\newcommand{\XMM}{{\it XMM-Newton~}}

\newcommand{\fermi}{{\it Fermi~}}

\newcommand{\ergs}{erg s$^{-1}$}

\newcommand{\ergsc}{erg cm$^{-2}$ s$^{-1}$}

\newcommand{\dg}{$^\circ$}

\newcommand{\lr}{$L_{5GHz}$}

\newcommand{\lx}{$L_{2keV}$}

\newcommand{\lgr}{$L_{100MeV-10GeV}$}

\newcommand{\Log}{{\rm Log\,}}

\newcommand{\er}{$\pm$}

\shorttitle{}

\shortauthors{Migliori G. et al.}

\begin{document}

\title{Jet Emission in Young Radio Sources: a \fermi-LAT Gamma-ray View}

\author{G. Migliori\altaffilmark{1,2}, A. Siemiginowska\altaffilmark{1}, B.C. Kelly\altaffilmark{3}, {\L}. Stawarz\altaffilmark{4,5}, A. Celotti\altaffilmark{6,7,8}, M.C. Begelman\altaffilmark{9}}

\altaffiltext{1}{Harvard-Smithsonian Center for Astrophysics, 60 Garden St., Cambridge, MA, 02138, USA}
\altaffiltext{2}{email: migliori@cfa.harvard.edu}
\altaffiltext{3}{Department of Physics, Broida Hall, University of California, Santa Barbara, CA, 93107, USA}
\altaffiltext{4}{Institute of Space and Astronautical Science, JAXA, 3-1-1 Yoshinodai, Chuo-ku, Sagamihara, Kanagawa 252-5210, Japan}
\altaffiltext{5}{Astronomical Observatory, Jagiellonian University, 30-244 Krak\'ow, Poland}
\altaffiltext{6}{Scuola Internazionale Superiore di Studi Avanzati (SISSA), via Bonomea, 265 - 34136 Trieste Italy}
\altaffiltext{7}{INAF-Osservatorio Astronomico di Brera, via E. Bianchi 46, 23807 Merate, Italy}
\altaffiltext{8}{INFN- Sezione di Trieste, via Valerio 2, 34127, Trieste, Italy}
\altaffiltext{9}{JILA, University of Colorado and National Institute of Standards and Technology, 440 UCB, Boulder, CO 80309-0440, USA}

\begin{abstract}
\noindent
We investigate the contribution of the beamed jet component to the high energy emission in young and compact extragalactic radio sources, focusing for the first time on the $\gamma$-ray band. We derive predictions on the $\gamma$-ray luminosities associated with the relativistic jet assuming a leptonic radiative model. The high energy emission is  produced via Compton scattering by the relativistic electrons in a spherical region at the considered scales ($\lesssim$10 kpc). 
Simulations show a wide range of \g-ray luminosities, with intensities up to $\sim10^{46}-10^{48}$ \ergs~depending on the assumed jet parameters. We find a highly linear relation between  the simulated X-ray and \g-ray luminosities that can be used to select candidates for  a \g-ray detection.
We compare the simulated luminosity distributions in the radio, X-ray and \g-ray regimes  with observations for the largest sample of  X-ray detected young radio quasars. 
Our analysis of $\sim$4-year \fermi  Large Area Telescope (LAT) data  does not give any statistically significant detection.   
However, the majority of the model-predicted \g-ray fluxes for the sample are near or below the current {\it Fermi}-LAT flux threshold and compatible with the derived upper limits. 
Our study gives constraints on the minimum jet power ($L_{jet,kin}/L_{disk}>0.01$), on a potential jet contribution to the X-ray emission in the most compact sources ($\lesssim1$ kpc) and on the particles to magnetic field energy density ratio in broad agreement with equipartition assumption.
\end{abstract}
\keywords{Galaxies: Active, Galaxies: Jets, Radiation mechanisms: non-thermal, $\gamma$-Rays: Galaxies}

\section{Introduction}
Young and compact radio sources represent the first stage in the evolution
of large extragalactic radio sources and constitute an important
fraction ($\sim 10\%-30\%$) of the radio source population \citep[see, however,][ for problems in selecting
genuine young sources against blazars]{Sta05,Tin05}.   
There are still open questions on the radio source initial phase, concerning its impact on the host galaxy and the evolutionary path from  compact to large (and giant) radio structures. \\
\indent 
Simulations indicate that during the first phase of the
expansion (which lasts $\approx$10$^5$ yrs) a powerful radio source
can interact with the interstellar medium (ISM) of its
host galaxy, and later on with  the intergalactic medium (IGM) \citep[][ and
references therein]{BS06,Wag12}.  Observations at radio and optical frequencies reveal signatures of
these interactions in nearby radio galaxies
\citep[e.g.][]{Tad07,Hol08,Lab08,Holt11,Sie12,Liu13}.  Both the mechanical and radiative energies released by the young
radio source can be relatively high and may have an important impact on the
host galaxy (e.g. influencing the star formation processes, affecting the amount and temperature of the gaseous fraction of ISM and IGM). Measurements of the radio source total power during the initial phase are thus important for unveiling feedback processes at work in the evolving galaxies.
\\
According to the current understanding, compact GigaHertz-Peaked Spectrum (GPS) radio sources evolve into Compact Steep Spectrum (CSS) sources\footnote{GPS and CSS radio sources are characterized by a linear size  $\lesssim$1 and $\lesssim$10-15 kpc, respectively, and a power at 1.4 GHz $P_{1.4\,GHz}\geq 10^{32}$ erg s$^{-1}$ Hz$^{-1}$. They typically display convex radio spectra with turn over frequencies between $\sim$0.1 and $\sim$1 GHz \citep[see][for a review]{Ode98}.} and finally become large-scale ($\gtrsim$20 kpc) radio sources \citep{Fan95,Beg96,Sne00}. The presence of a class of even more compact and young radio sources, so-called High Frequency Peakers (HFP), is still under investigation \citep[][for a review]{Dal00,Ori09}.
The high fraction of young sources in radio samples (in comparison with the large-scale ones) suggests a more complex evolutionary picture: intermittency of the radio activity \citep{RB97} or the presence of a population of short-lived sources \citep[never reaching the large-scale stage, e.g.][]{Ale00,Kun10,OMD10} are possible explanations. Furthermore, it is not clear which morphology GPS and CSS will finally display, i.e. whether they will evolve into Fanaroff-Riley type I (FRI) or type II (FRII) sources \citep{FR74}.\\
\indent
The high-energy domain may provide important clues on the most energetic processes associated with the young source phase. So far, most of the high-energy investigations have been done in the X-rays (0.5-10 keV) using \Cha and \XMM observations
\citep{Sie08,Ten09}. These show that young sources  are relatively X-ray luminous ($L_X\sim10^{42}-10^{46}$ \ergs).
However, the entire radio structure is typically enclosed within angular scales which cannot be resolved by X-ray telescopes (a few arcseconds or smaller), thus the X-ray morphology is usually not accessible \citep{Sie09}.   
Studies of the origin of the X-ray emission, including an identification of distinct high-energy emission components, rely mainly on the analysis of X-ray
spectral features, which is hampered by the limited photon statistics \citep{Gua06,Vin06,Sie08,Ten09}.
As a consequence, it is hard to disentangle the contribution of the jet- and lobe-related
non--thermal emission from the one related to the accretion processes
(either the X-ray thermal emission of the innermost part of the disk
or that resulting
from Comptonization of disk photons by electrons in a hot corona).\\
\indent 
A significant level of non-thermal X-ray flux can be
produced via Inverse Compton (IC) scattering of different seed photons in the
compact lobes of GPS radio galaxies
\citep{Sta08}. Given the typical GPS linear sizes ($\lesssim$1 kpc)
the nuclear photon fields, e.g.  optical-UV disk and IR torus photons,
are intense enough to provide, when Comptonized, X-ray luminosities of
the order of $10^{44}$--$10^{46}$~erg~s$^{-1}$ \citep{Sta08,Ost10}.
In the case of GPS and CSS quasars, the
closer jet alignment to the line of sight should favor the beamed jet
component. This is certainly observed in the radio band \citep{Fan90},
while the X-ray behavior remains more elusive. 
The jet X-ray emission can be important depending on its parameters
(inclination, bulk motion, intrinsic power).
In the case of the CSS
quasar 3C~186, we have shown that the jet could contribute to the
total X-ray emission when it develops a
complex velocity structure \citep[see][]{Mig12}.\\
\indent   
The $\gamma$-ray band can be useful to study many aspects of the physics of young radio sources.
At these frequencies the disk-corona
component is expected to rapidly drop in intensity, while both beamed
and unbeamed non-thermal emission could be still strong and detectable
in powerful sources. The intensity of the $\gamma$-ray flux depends on
several factors. For the case of compact symmetric objects (CSO), the
model presented in \citet{Sta08} predicts a significant isotropic
non-thermal emission in the 0.1 -- 100 GeV band produced in the radio
lobes, however within a wide range from $\sim10^{41}$ \ergs~up to $\sim10^{46}$ \ergs~ depending on the source parameters. In this scenario, the non-thermal high-energy emission may even
dominate the total radiative output of the source, and as such it is crucial for
constraining the total energetics of the expanding source.  
\citet{Ito11} and \citet{Kin13} have shown that non-thermal IC emission may arise from a shell of shocked ISM driven by the expanding radio source. The estimated GeV and TeV emission from the mini shells associated with powerful (i.e. with jet powers of the order of $\sim10^{46}$ \ergs) and compact ($\sim$5 pc) sources  could be potentially detectable by the next generation of Cherenkov telescopes. The \g-ray emission related to the hadronic component within the compact lobes of CSO was discussed in \citet{KA11}.  
 \citet{Kin09} predict bright GeV luminosities produced via bremsstrahlung in the cocoons and radio lobes of CSO. In this case, a \g-ray detection with \fermi-LAT would be possible for nearby ($\lesssim 10^2$ Mpc) sources if the jet is powerful and made of e$^{\pm}$ plasma.\\
\indent 
A dedicated study of the jet component in young radio sources and its possible \g-ray emission has not yet been performed. 
We have therefore investigated for the first time the \g-ray properties of a sample of young
radio sources using the Large Area Telescope (LAT) onboard the \fermi satellite \citep[see][]{Atw09}.  The goals of the study are:
1- to refine model predictions for the \g-ray emission of the jets of young radio sources; 2- to constrain the entire non-thermal
high-energy emission continua of these objects. 
We adopt the
following approach:  we first set up a general leptonic synchrotron and inverse
Compton (IC) model for the non-thermal emission of
jets in compact radio sources; then we generate a library of their broad-band spectral energy distributions (SED)
for a wide range of selected jet parameters ( linear
sizes, jet powers, jet bulk velocities etc.). 
 While previous studies focused on the isotropic emission associated with the extended structures of young sources, i.e. lobes and cocoons, in this work we concentrate on the jet beamed radiation and consider powerful radio quasars. Indeed, an important difference between jet and lobe emission arises from the jet relativistic (bulk) motion: the jet emission is
relativistically beamed and its detection also depends on the observer viewing angle.
In models presented in the literature \citep[see e.g.][]{Sta08,Kin09,Ito11,KA11,Kin13} the emission of the source is consistently calculated by coupling the evolution of the electron distribution to the source dynamical expansion. Here, rather than following the evolution of a single source, we simulate the emission for a distribution of sources assuming a wide range of values for the jet parameters.
These simulations are compared with the analyzed \fermi-LAT
data for the considered GPS and CSS sample. We discuss the results of the LAT analysis on the sample and the
implications for non-thermal models.

\section{Jet Model}
We set up a model to evaluate the non-thermal (beamed)
emission from jets in young radio sources. 
The jet emission is computed using a general
leptonic radiative model. 
In the following, we describe the main features of the model.\\
\indent
We assume a simple geometry for the emission region (see Figure \ref{f1} for a schematic illustration). The bulk of the jet emission is produced in a spherical
knot, located at a distance $z_d$ from the central Black Hole (BH). 
The knot  is
  moving with a bulk Lorentz factor $\Gamma$. 
The electrons in the knot radiate via synchrotron and IC emission. Seed photons for
  the IC mechanism are the synchrotron emission of the knot
   and external radiation fields (synchrotron-self-Compton, SSC, and external Compton, EC, processes, respectively).  
The latter are UV photons from the accretion disk \citep[EC/disk][]{DSM92} or disk emission
  reprocessed by the dust surrounding the central engine and emitted at IR wavelengths \citep[EC/dust][]{Bla00,Sik09}.
The model also takes into account the high energy emission that may be associated with the presence of a longitudinal velocity
  structure inside the jet. This is done by adding a radiative
  field of  synchrotron photons (EC/syn) produced outside the knot in a blazar-like component
  located at the base of the jet, moving with a highly
  relativistic speed $\Gamma_{bl}$ \citep[see][]{Cel01,Mig12}.  
The intensity of EC/syn luminosity can be comparable to the high energy emission potentially produced by the blazar core itself (e.g. via SSC or Compton scattering of the photons of the Broad Line Regions, BLR, and of the dusty torus). Therefore  we included the EC/syn contribution to the total jet high-energy emission only if it is larger than the blazar one for a given parameter set. The EC luminosity is treated following prescriptions for anisotropic radiation fields in the reference frame of the emitting region \citep[see][]{DSM92,Sik94,Der95,SSO03}.\\ 
\indent
In order to reduce the number of free parameters in the model, we linked together or fixed some of the quantities (see Table \ref{t1}). The height of the emitting knot $z_d$ is equal to the jet linear size, measured from the core to the jet termination (hereafter LS). The jet has a conical geometry with a semi-aperture angle equal to 0.1 (R=0.1LS), which allows for reasonable sizes  of the knot radius R (1 pc to 1 kpc) within the considered linear scales (LS$\leq$10 kpc). \\ 
\indent
The fraction of disk
photons reprocessed by nuclear
dust in a putative torus is fixed to $L_{dust}=0.1 L_{disk}$, in agreement with luminosities found for type 1 quasars \citep{Hat08}, where the disk emission can be directly observed.
Since we are not aiming at a detailed model of the source spectrum,
thermal emission from the disk and the dust is simply modeled with a blackbody spectrum. 
We set
the disk temperature $T_{disk}$ to $\sim 3\times10^4$ K. The dust temperature ($T_{dust}\sim$370 K) is derived from the observed 
peak frequency of the IR emission \citep[$\nu_{dust}=3\times10^{13}$ Hz][]{Clea07} using the formula
$T_{dust}=h\nu_{dust}/(3.93k)$ \citep[see][]{GT09}. The considered spatial scales are larger than 1 pc, thus we always treat the disk
as a point-like source of photons, with the energy density\footnote{Hereafter, primed energy densities and luminosities are in the jet knot comoving frame.} of the disk photons, $U'_{disk}$, scaling as $1/LS^{2}$ (and $1/\Gamma^{2}$). For simplicity, we assumed a
thin spherical shell for the dust spatial distribution, the shell radius being 
$R_{dust}=2.5\times10^{18}\sqrt{L_{disk}/10^{45}}$ cm \citep[][]{Bla00,Sik02}.
The dust photon energy density, $U'_{dust}$, is computed using Equation (21) in \citet{GT09} when  $z_d<R_{dust}$ and considering a point-like source of photons when $z_d>R_{dust}$. The value of $U'_{dust}$ at the discontinuity point, $z_d=R_{dust}$, is the average between the dust energy densities immediately inside and outside the shell. \\
\indent
We assume a link between the accretion and ejection processes related to the supermassive BH. The power carried by the jet in radiating particles and Poynting flux ($L_{jet,kin}$) is  proportional to the disk luminosity, $L_{disk}$. A relation between the jet and disk powers in extragalactic radio sources has been considered by \citet{RS91}, who found that the average jet kinetic power, necessary to support the radio lobes, is comparable with the accretion disk luminosity. In powerful blazars, an energetically dominant proton component is required to account for the radiative jet luminosities, and the jet total power can be even larger than the accretion disk luminosity \citep[$L_{jet,kin}/L_{disk}=1.0-10$][]{CG08,Ghi10}.  
In GPS radio galaxies, \citet{Ost10} inferred $L_{jet,kin}/L_{disk}=0.01-0.1$ based on their modeling of the lobe emission.\\
\indent
We set the parameters of the blazar-like blob referring to the estimated ranges of the sample of Flat Spectrum Radio Quasars (FSRQ) \citep{CG08}: the blazar blob emits 10\% of the jet kinetic power via synchrotron emission and has a bulk motion of $\Gamma_{bl}=10$.\\ 
\indent
To summarize, in the model $LS$, $L_{disk}$, $\Gamma$ and the observer viewing angle $\theta$ are free input parameters, $z_d$ is equal to $LS$, and the jet kinetic power $L_{jet,kin}$ and the dust emission $L_{dust}$ are both proportional to $L_{disk}$.  
In addition, we assume a moderate dominance of particles over magnetic field energy density in the knot: $U'_e/U'_B=10$, $U'_e$ being the energy density of the radiating electrons and $U'_B$ the magnetic field energy density, respectively. Observations indicate that radio lobes in young sources are in equipartition conditions \citep[see][]{OD08}. However, in jet knots of FRII radio galaxies, radio and X-ray observations point to intensities of the magnetic field below the values expected under the assumption of energy equipartition with the radiating particles, when the X-ray emission is ascribed to IC emission \citep[see e.g.][]{KS05}. Similarly, SED modeling of multiwavelength emission of FSRQ and BL Lac objects points to particle dominated jets \citep{CG08}.\\ 
\indent
We allow the shape of the electron energy distribution (hereafter EED) to be either a simple power law, $N(\gamma)=k_e \gamma^{-p}$, or a broken power law.
The minimum and maximum electron Lorentz factor,
$\gamma_{min}$ and $\gamma_{max}$, are fixed to 10 and 10$^5$, respectively. 
The parameter values of the EED ($p=2.68$ for the single power law spectral index; $p=2.1$ and $p2=4.0$ and $\gamma_{break}\sim2\times10^3$ for the spectral indexes of the broken power law and the energy break, respectively)  are derived from observations as discussed in the following section.  
 The model
parameters (and with the assumed values) are listed in Table \ref{t1}.\\

\section{Simulated Jet SED}
\begin{figure}
\centering
\includegraphics[scale=0.45]{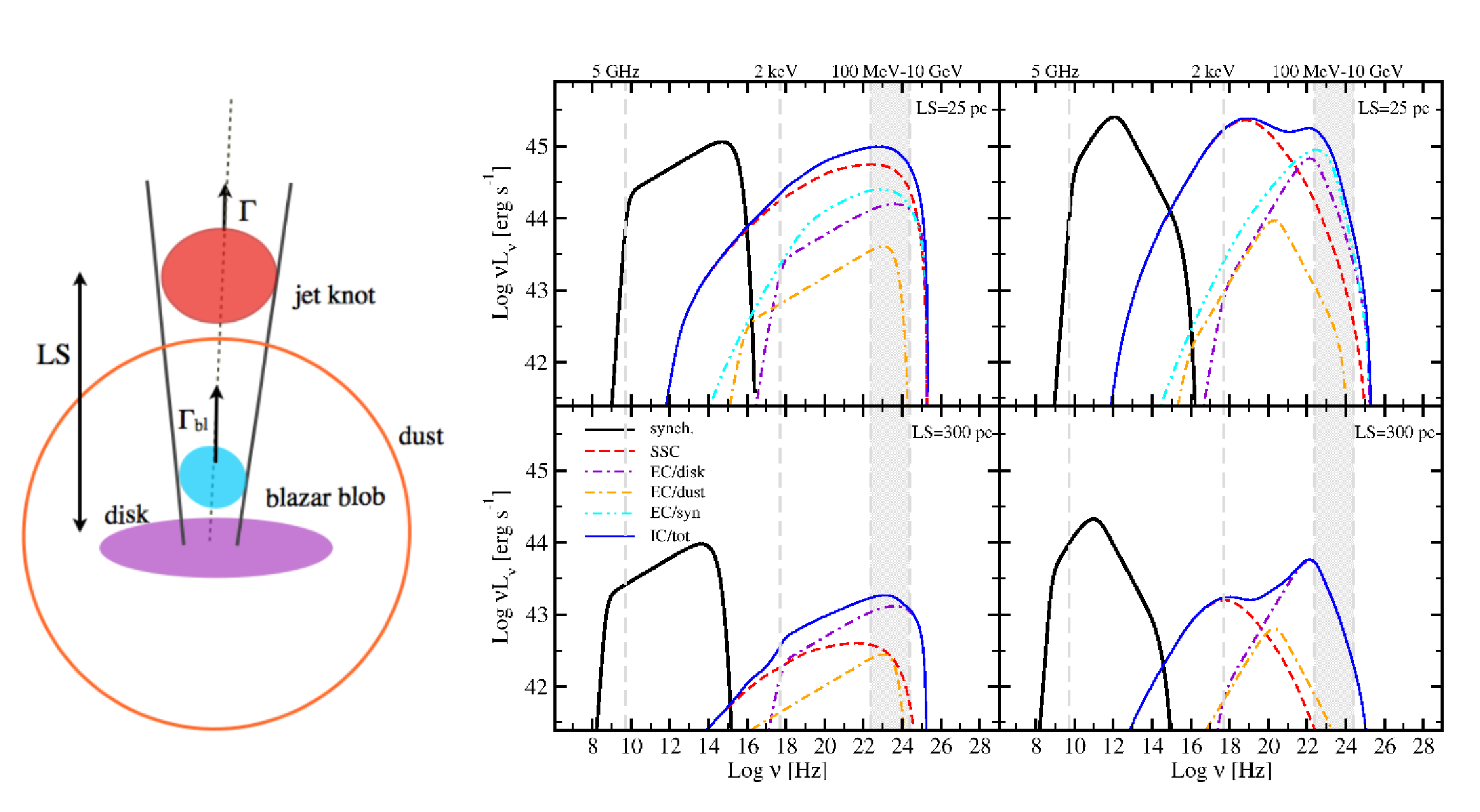}
\caption{Left: sketch of the jet model: the observed jet luminosities are produced in a jet spherical knot (in red). The external photon fields contributing to the jet EC emission are schematically illustrated: disk, dust and blazar component at the base of the jet (in cyan). Right: Simulated SED for a 25 pc jet  (upper panels) and a 300 pc jet (lower panels). The jet parameters are: $L_{jet,kin}=L_{disk}=10^{46}$ \ergs, $\Gamma=2.0$, $\theta=30^\circ$. Left panels: the assumed EED is a simple power law with $\gamma_{min}=10$, $\gamma_{max}=10^5$, p=2.68. Right panels: the assumed EED is a broken power law with $\gamma_{min}=10$, $\gamma_{max}=10^5$, $\gamma_{break}=2\times10^3$, p=2.1, p2=4.0. Note that the EC/syn component is not included in the SED of the 300 pc jet (see text).}
\label{f1}
\end{figure}

We simulate the jets' SED for a range of LS  and bolometric disk luminosities, $L_{disk}$.
Our simulations cover the range of LS and $L_{disk}$ values of the sample of young radio quasars with an X-ray detection presented in \citet{Sie08}, $LS=0.01-10$ kpc and $L_{disk }= 10^{45}-10^{47}$ \ergs~respectively. In fact, we anticipate that X-rays provide important constraints on the model predicted $\gamma$-ray luminosities.
The optical-UV emission of the GPS and CSS quasars displays the typical features of broad-line quasars, with detected UV-bump and broad emission lines, and the measured intrinsic absorption columns in the X-rays are less than a few $\times 10^{21}$ cm$^{-2}$ \citep{Sie08}. Thus, we do not expect the disk emission to be heavily affected by obscuration. Since the observed disk emission is not modified by other orientation effects, e.g. relativistic beaming, we use it to define the luminosity range of our simulations.\\ 
\indent
Mildly to highly relativistic jet velocities are considered, with values of the bulk motion $\Gamma$ between 1.4 to 10, where the minimum value refers to the study presented in \citet{MH09}.  We allowed observer's viewing angles, $\theta$, in the 10\dg~to 50\dg~range so that we exclude the most closely aligned objects (i.e. blazar sources).  Inclination angles $\lesssim$10\dg~would also imply larger jet linear sizes for the quasars in the sample than typical GPS and CSS sources. It is however true that single components within the jet (i.e. a single knot or the inner jet) might have smaller $\theta$.\\ 
\indent
The shape of the EED in young sources is not well known. Here, we consider a single and a broken power law. 
The spectral index for the single power law EED is set to the median X-ray photon index ($\Gamma_X=1.84\pm0.24$, $p=2\Gamma_X-1$) of the sample of GPS and CSS quasars \citep{Sie08}. 
In the broken power law EED, we set the spectral index below and above
the energy break to $p$=2.1 and $p2$=4.0, respectively. The low-energy index is consistent (within the errors) with the average $\Gamma_X$ in radio loud quasars \citep[$\Gamma_X=1.57\pm0.08$ and $\Gamma_X=1.55\pm0.17$,][respectively]{Bec94,Bel06}.  
We refer to \g-ray observations of misaligned radio sources to derive the spectral index above the energy break: we consider the sample of  misaligned radio sources detected by {\it Fermi}-LAT and select the central value in the range of their measured \g-ray photon indexes \citep[$\Gamma_\gamma \sim 1.9-3.0$][]{Aab10, Kat11, Ack12}.
The energy break is fixed to an intermediate value between the minimum and maximum electron Lorentz factor, $\gamma_{break}\sim2\times10^3$. 
Similar EED functions have been inferred by modeling the broad band SED of hot spots \citep[see the cases of Cygnus A, PKS 1421-490 and 3C 445 in][respectively]{Sta07,God09,Per10} and blazar jets \citep{CG08,GT09,Sik09}. 
In this work, we do not discuss a specific theoretical scenario for the particle acceleration process. However, it is interesting to note that, in jets, indications for electron injection spectra that deviate from the standard E$^{-2}$ form (related to first-order Fermi acceleration process in non-relativistic shocks) come from observations \citep{Sta07,God09,Per10,CG08,GT09,Sik09} and simulations \citep[see e.g. the results of particle in cell simulations presented in][and references therein]{SS11}.\\
\indent
We allow three jet to disk power ratios, $L_{jet,kin}/L_{disk}=$0.01, 0.1 and 1.0, hence, given the assumed $L_{disk}$ values, the minimum and maximum jet powers in our simulations are $10^{43}$ and $10^{47}$ \ergs, respectively. These values cover the ratios previously inferred for radio sources, including GPS radio galaxies \citep[e.g.][]{Ost10}. We do not consider the case of $L_{jet,kin}/L_{disk}>1$, which was derived for FSRQ and could be possibly associated with flaring states limited in time \citep[see][and references therein for a discussion]{CG08,Aab10a,Tan11}.   \\
\indent
 In Figure \ref{f1} we show the simulated SED for a
jet with $L_{jet,kin}=L_{disk}=10^{46}$ \ergs, moving with $\Gamma=2.0$ and observed at
$\theta$=30\dg. The SED are for two considered linear sizes, LS=25 pc (upper panels) and LS=300 pc (lower panels) and for the two assumed functions of the EED (the single power law in the left panels, and the broken power law in right panels).\\ 
\indent
The low energy part of the spectrum includes synchrotron self absorption effects. 
The characteristic convex spectral shape observed at radio frequencies in young sources could also be due to free-free absorption by an inhomogeneous ambient medium.
However, such effect is likely to be more important in the radio lobes' synchrotron spectrum \citep[see][]{Bic97,Beg99,Sta08}.\\
\indent
For the single power law EED, the SED of the 25 pc jet above $\sim$10$^{16}$\,Hz is dominated by one mechanism, i.e. SSC radiation. For the same linear size, in the broken power law case the X-ray luminosity (and up to $\approx$10$^{21}$ Hz) is mainly produced via SSC while the \g-ray luminosity ($\sim$10$^{22}$-10$^{24}$\,Hz) via EC/syn. 
The energy density of the synchrotron photons produced by the blazar-like component scales as $\Gamma_{bl}^4/\Gamma^2 \times1/z_d^2$ \citep[in the knot rest frame, see][]{Cel01,Mig12}. Therefore, the EC/syn luminosity is expected to be intense in powerful jets that decelerate on {\bf scales of} tens of parsecs. In misaligned sources (e.g. $\gtrsim$20\dg~for our model parameters), such component can be brighter than the high-energy emission from the inner blazar component.\\
\indent
The underlying EED shape determines the observed X-ray to \g-ray spectrum, with a slope of the SED (for the 25 pc jet) that is steep for the single power law EED and flat for the broken power law. The difference between the two SED holds true even when the EC/syn component is not present. In fact, the EC/disk luminosity peaks in the \g-ray band and still dominates over the SSC \g-ray emission in the broken power law EED case. \\ 
\indent
The radiative output of the jet decreases at all wavelengths for increasing LS. Furthermore, the dominant contributions to the X-ray and \g-ray bands may change as the source expands. Figure \ref{f1} shows that the EC/disk luminosity  dominates the high-energy output of the 300 pc jet. This follows from the model scaling relations for the synchrotron and IC luminosities. Given the jet kinetic power in radiating particles and magnetic field:\\ 
\begin{equation}
L_{jet,kin}\approx \pi R^2 c \Gamma^2 (U'_e+U'_B) \approx 10^{-3} \pi c LS^2 \Gamma^2 U'_e,\\
\label{eq1}
\end{equation}
where $U'_e=10 U'_B$ and $R=0.1 LS$, the total synchrotron luminosity in the jet comoving frame depends on the jet power and the volume of the emitting region:\\
\begin{equation} 
L'_{syn}\propto U'_e U'_B V \propto L_{jet,kin}^2 LS^{-1} \Gamma^{-4}.
\label{eq2}
\end{equation}
The EC luminosities related to the disk emission (EC/disk, EC/dust) scale in the same way (for $z_d=LS$):\\
\begin{equation}
L'_{EC}\propto U'_e U'_{disk/dust} V \propto L_{jet,kin}^2 LS^{-1} \Gamma^{-4}, 
\label{eq3}
\end{equation}
for $L_{jet,kin}\propto L_{disk}$ and $z_d>R_{dust}$, i.e. when the emitting knot is out of the torus region (for $z_d\lesssim R_{dust}$: $L'_{EC/dust}\propto \Gamma^0$), while the SSC luminosity scales as:
\begin{equation}
L'_{SSC}\propto U'_e U'_{syn} V \propto L_{jet,kin}^3 LS^{-2} \Gamma^{-6},
\label{eq4}
\end{equation}
where $U'_{syn}$ is the energy density of the synchrotron photons in the jet/knot rest frame.\\
\indent
The SSC luminosity decreases faster than EC luminosities with LS. This is evident in the SED for the 300 pc jet: for both assumed EED the EC/disk luminosity peak is higher than the SSC one. Interestingly, in the broken power law EED this also changes the X-to-\g-ray luminosity ratio (and slope) with respect to the SED of the 25 pc jet.
Note that the EC/syn component is not considered at LS=300 pc, as it becomes fainter than the blazar emission. 

\subsection{Simulated Gamma-ray Luminosity Distributions}
\begin{figure}
\centering
\includegraphics[scale=0.33, angle=-90]{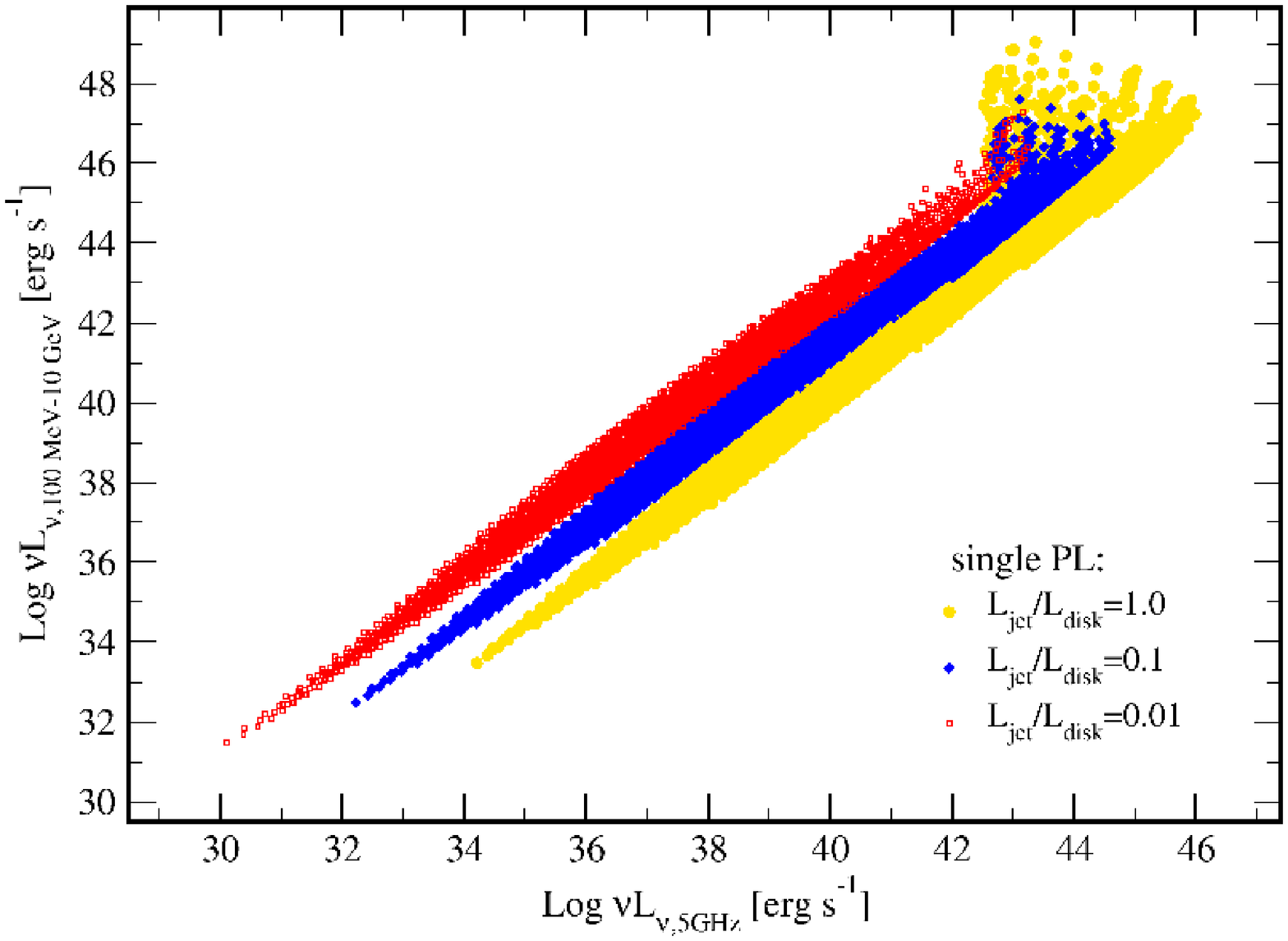}
%\vspace{0.3cm}
%\hspace{0.2cm}
\includegraphics[scale=0.33, angle=-90]{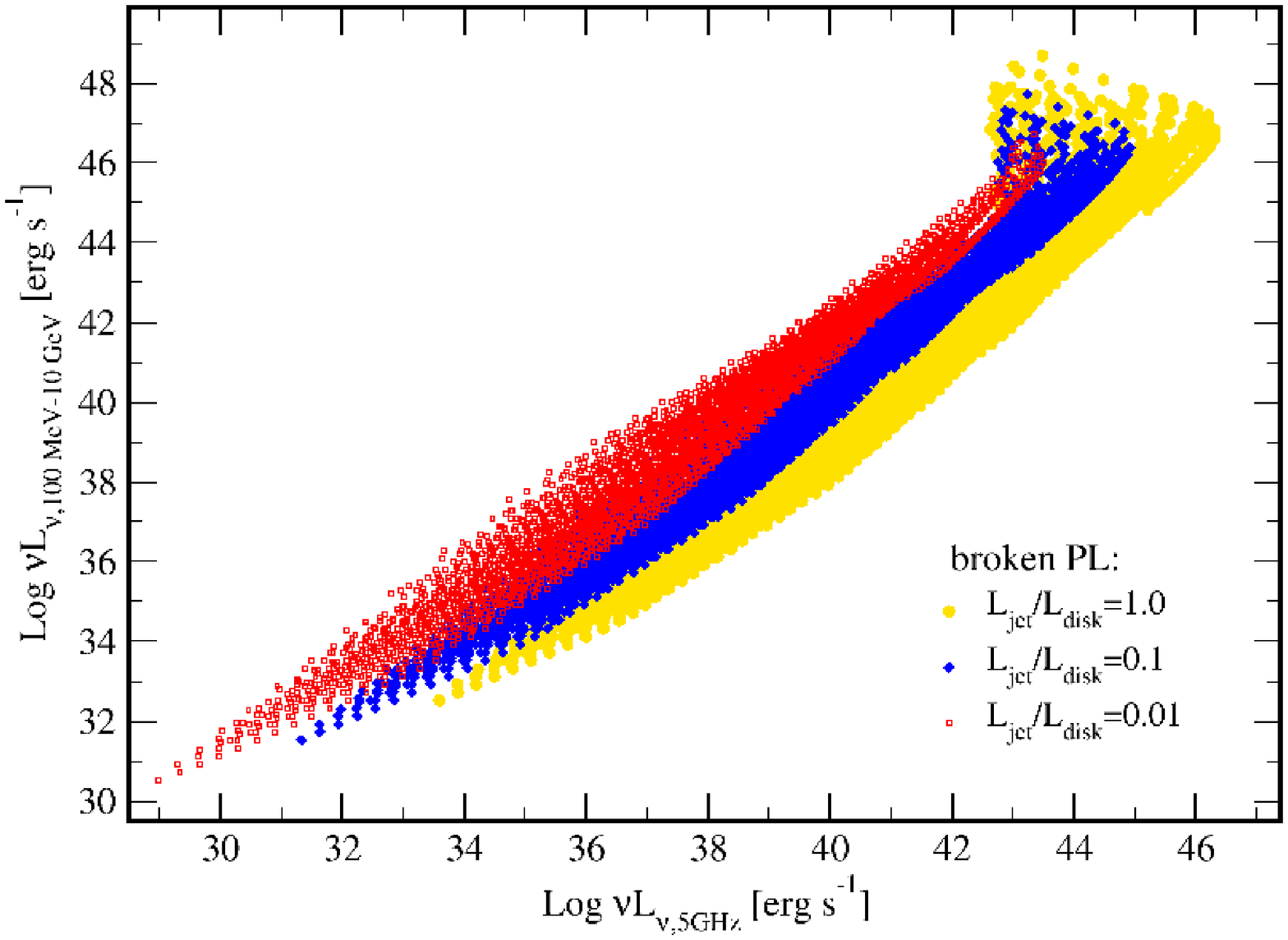}
%\vspace{0.3cm}
\includegraphics[scale=0.33, angle=-90]{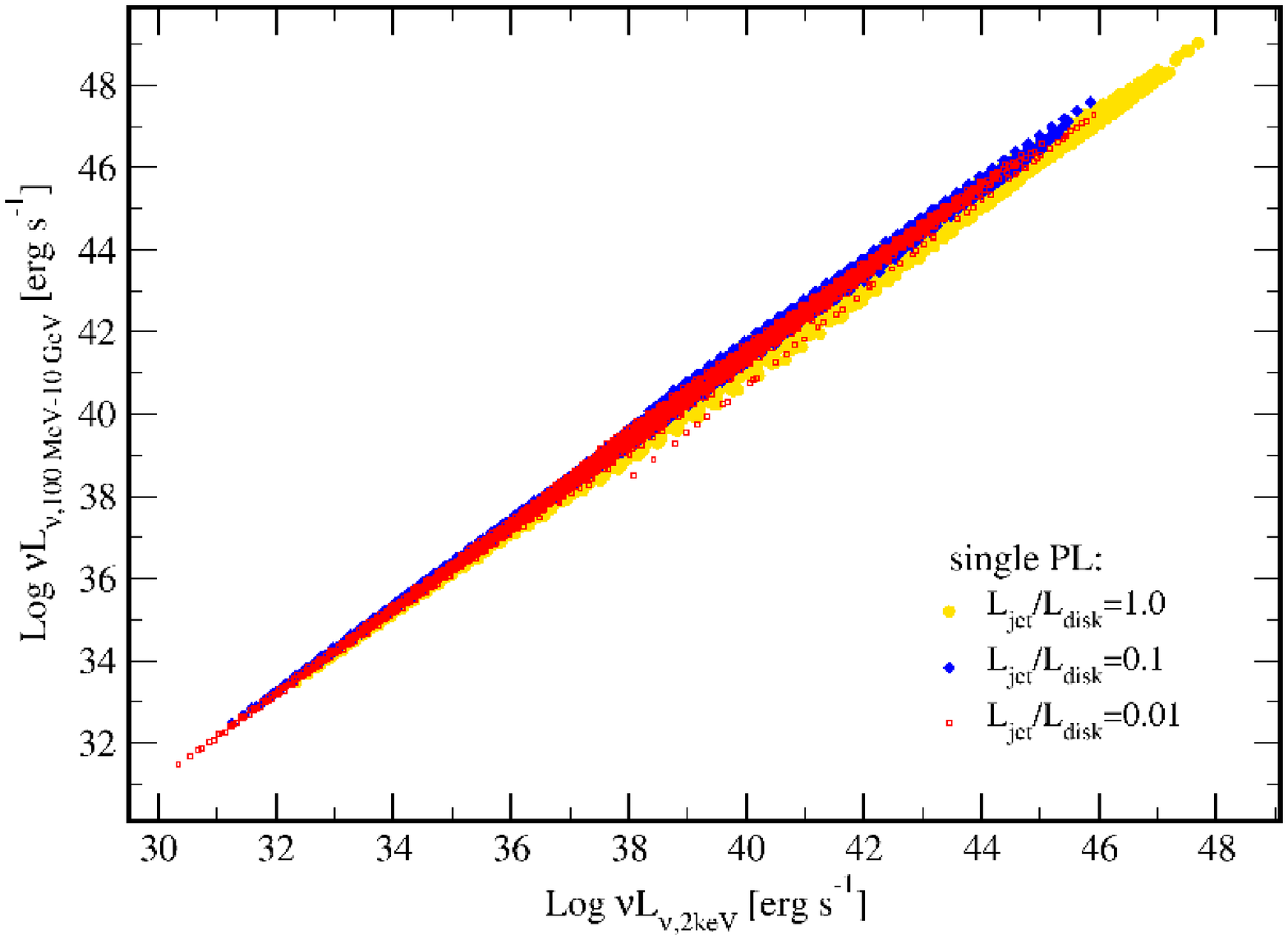}
%\hspace{0.2cm}
\includegraphics[scale=0.33, angle=-90]{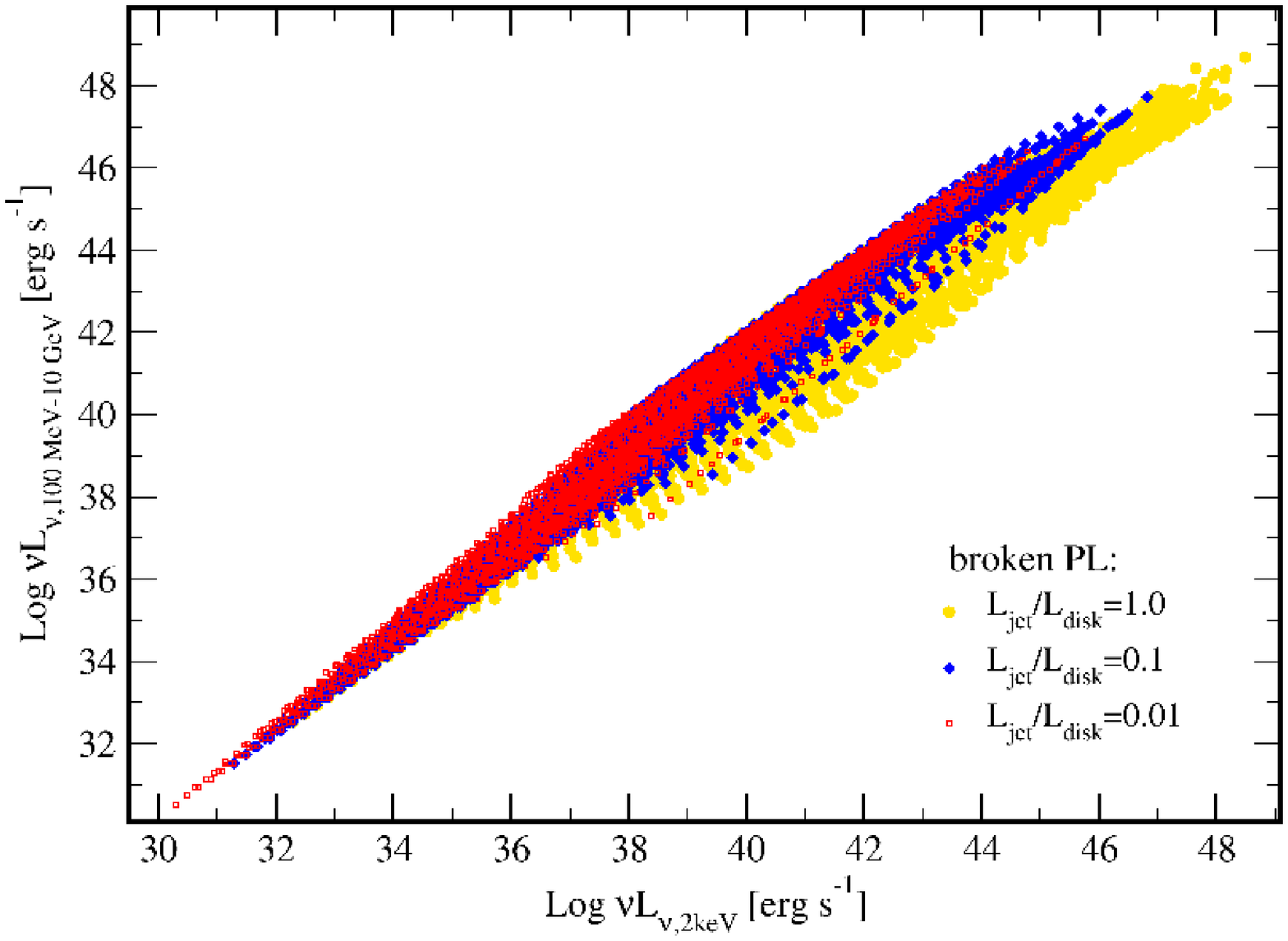}
\caption{Gamma-ray (100 MeV - 10 GeV) luminosities are plotted as a
  function of the radio (5 GHz, upper panels)
  and X-ray (2 keV, lower panels)  luminosities for the simulated jets. All luminosities are calculated in the observer's reference frame. The simulated jets cover a range of linear sizes (LS= 10 pc
  - 10 kpc), bulk Lorentz factors ($\Gamma=$1.4-10) and observer
  viewing angles ($\theta=$10\dg-50\dg). Yellow solid circles are the simulated sources assuming $L_{jet,kin}=L_{disk}$, the blue solid diamonds the simulated sources for $L_{jet,kin}=0.1 L_{disk}$ and the red empty squares for $L_{jet,kin}=0.01 L_{disk}$. The (bolometric) disk luminosities $L_{disk}$ are in the range $10^{45}-10^{47}$ \ergs.
Left panels show simulations for
  a single power-law EED with $p=2.68$ and right panel for a
  broken power law with $p=2.1$, $p2=4.0$ and
  $\gamma_{break}=2\times10^3$. In the EED, the minimum and
  maximum random Lorentz factors are $\gamma_{min}=10$ and
  $\gamma_{max}=10^5$ respectively. }
\label{f2}
\end{figure}

\begin{figure}
\centering
\includegraphics[scale=0.55, angle=-90]{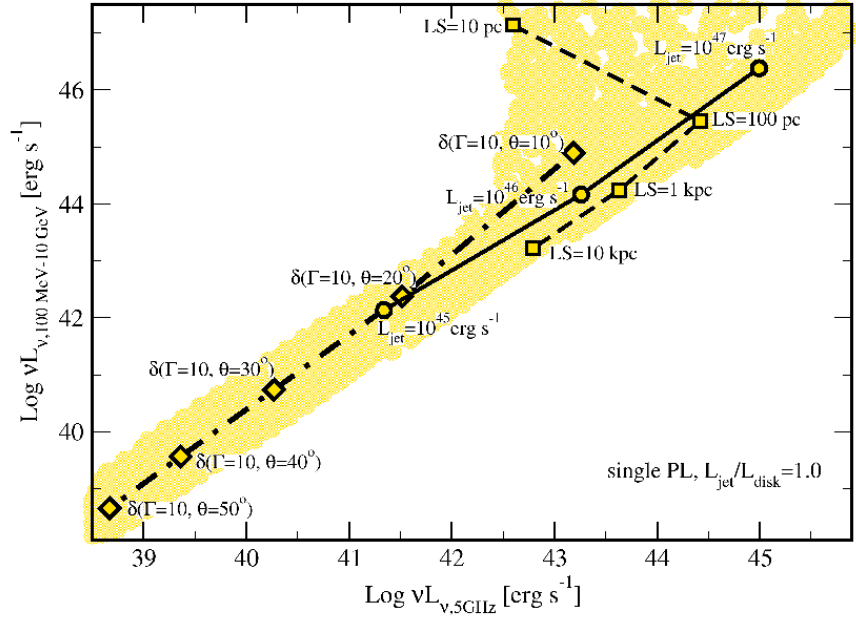}
\caption{Dependencies of the simulated distribution (for the case of $L_{jet,kin}=L_{disk}$ and a single power law EED) on the free parameters are highlighted. Each trail shows how the position of a source changes in the \lr-\lgr~plot for a single varying parameter. The solid black line connecting the circles shows the evolution as a function of the jet power for a source with LS=100 pc, $\Gamma=4$ and $\theta=20$\dg. The dashed line and squares show the evolution with LS for a source with $L_{jet,kin}=10^{46}$ \ergs, $\Gamma=1.4$ and $\theta=20$\dg. The dot-dashed line and diamonds show the effect of beaming factor (varying $\theta$ and leaving $\Gamma$ fixed) for a source with  $L_{jet,kin}=10^{46}$ \ergs~and $LS$=10 pc.}
\label{f2a}
\end{figure}

 For each of the simulated SED we computed the synchrotron
luminosity at 5 GHz (\lr), the X-ray luminosity
at 2 keV (\lx)
and integrated 100 MeV - 10 GeV luminosity (\lgr) in the observer rest frame. 
The 2 keV and 100 MeV-10 GeV luminosities are obtained by adding up all the IC contributions (SSC, EC/disk, EC/dust and EC/syn) at each frequency. For the assumed jet parameters' ranges, the synchrotron emission never significantly contributes at 2 keV. 
We recall that the external Compton emission on the
blazar-like synchrotron photons (EC/syn) is included only when, for a
given parameter set, it is larger than the estimated high-energy emission of the blazar component itself, assuming for the latter SSC and IC on the broad line and dusty torus photons. This happens in $\sim$10\% of the cases for the simulated distributions.\\
\indent
Figure \ref{f2} shows the \g-ray luminosity of the simulated jets as a function of their radio (\lr--\lgr, upper panels)  and X-ray luminosities (\lx--\lgr, lower panels), for the two considered EED (left and right panels for the single and the broken power law EED, respectively) and jet to disk power ratios ($L_{jet,kin}/L_{disk}=$0.01, 0.1 and 1.0). \\
\indent
Figure \ref{f2a} illustrates how the input model parameters, $L_{jet,kin}$, LS, $\Gamma$ and $\theta$, shape the simulated \lr-\lgr~distributions. We use the $L_{jet,kin}/L_{disk}=$1.0 distribution  (in the single power law EED case) as an example: each trail in the figure highlights the path in the luminosity-luminosity space when varying a single input parameter while keeping the others fixed.
Powerful ($L_{jet,kin}\sim10^{47}$ \ergs) and small (LS$\sim$10 pc) sources are the most intrinsically luminous, while low power and large (LS$\sim$10 kpc) sources are the least intrinsically luminous, as expected from the scalings of the luminosities in (\ref{eq2}), (\ref{eq3}), (\ref{eq4}).  
Effects related to the jet speed, orientation with respect to the line of sight, and to the selected monochromatic luminosities (in the observer rest frame) are also important. Modifications of the emitted luminosities in the observer rest frame due to $\theta$ and $\Gamma$ are expressed via the Doppler factor $\delta$ ($\delta=1/(\Gamma-\sqrt{\Gamma^2-1}\times \cos \theta)$). The intrinsic luminosity transforms in the observer frame as $\nu L_{\nu}\sim \nu L'_{\nu} \delta^4$, and the observed luminosity of a jet may change by orders of magnitude for increasing $\theta$ depending on its bulk motion. For $\Gamma=$10, for instance, $\delta$ changes from  $\sim$5 to $\sim$0.3 going from $\theta=$10\dg~to $\theta=$50\dg, with a correspondent variation of more than 4 orders of magnitude in \lr~and \lgr. The position of a  jet  (with a given set of $LS$, $L_{jet,kin}$ and $\Gamma$) can therefore migrate across the \lr\,(\lx)--\lgr~space depending on $\theta$ (see Figure \ref{f2a}). 
Doppler effects may be particularly relevant for monochromatic (or band) luminosities close to the extremes of the synchrotron and IC spectra, where even a small shift along the frequency axis ($\nu=\nu' \delta$, with $\nu'$ and $\nu$ the emitted and observed frequency, respectively) significantly changes the observed luminosity. \\
\indent 
The \lr-\lgr~distributions (and the \lr-\lx~distributions in Figure \ref{f3}) present a fan-like shape, with an increasing spread of the \lr-\lgr~relation for \lr$\gtrsim 10^{42}$ \ergs. 
In intrinsically powerful sources \lr~drastically increases when the jet (and therefore the volume of the emitting region) expands up to a few hundred parsecs and the 5 GHz frequency shifts from the optically thick to the optically thin part of the synchrotron spectrum.
For LS$\gtrsim$100 pc, \lr~progressively declines (following $L'_{syn}\propto LS^{-1}$) as shown in Figure \ref{f2a}.  
In low luminosity sources, the turn-over frequency from the optically thick to the optically thin regime is below 5 GHz at all the considered LS, thus \lr~simply declines $\propto LS^{-1}$ (when all the other input parameters are fixed). 
This explains the fact that for lower jet power intervals  (i.e. when $L_{jet,kin}/L_{disk}$=0.01 is assumed) the \lr-\lgr~distributions are more linear, though with dispersion.\\ 
\indent
For decreasing $L_{jet,kin}/L_{disk}$, the whole \lr-\lgr~distribution shifts to lower radio and $\gamma$-ray maximum and minimum luminosities. Interestingly, jets with $L_{jet,kin}/L_{disk}=0.01$ can still be relatively \g-ray luminous (\lgr$\sim10^{46}-10^{47}$ \ergs). 
We also note that, for a fixed \lr, the highest \lgr~ are obtained for the smallest jet-to-disk ratio ($L_{jet,kin}/L_{disk}=$0.01).
This is due to the fact that  EC/disk gives the dominant contribution in the \g-ray band.
Finally, the \lr-\lgr~relation shows a lower dispersion for the single power law EED than for the broken power law EED, as expected for a simpler spectral shape.\\
\indent
The simulated \lgr-\lx~distributions (Figure \ref{f2}, lower panels) are linear for all the considered $L_{jet,kin}/L_{disk}$, the X-ray and \g-ray emissions being produced by the same electron population via IC mechanism.
The distributions obtained from the broken power law EED have a larger dispersion than the simple power law EED. In fact, in the former case \lx~and \lgr~can be related to different IC emissions, e.g. SSC emission in the X-rays and EC in the \g-rays (as in the example in Figure \ref{f1}).   
A dominant SSC contribution in the X-rays is expected in powerful and small jets ($L_{jet,kin}\sim10^{46}-10^{47}$ \ergs, LS$\sim$10 pc), since $L'_{SSC}\propto L_{jet,kin}^{3} LS^{-2}$. For $L_{jet,kin}/L_{disk}=$0.01 and 0.1, the relative increase of EC/disk and EC/dust over SSC in the X-ray band reduces the spread of the \lx-\lgr~distribution with respect to the $L_{jet,kin}/L_{disk}=$1.0 case. In addition, in the \g-ray band attenuation of the IC flux may happen when the scattering occurs in the Klein-Nishina regime (which is included in the model computation of the IC spectrum).

\subsection{L$_{5GHz}$, L$_{2keV}$, L$_{100MeV-10GeV}$ relation}
 The simulated \lr--\lgr~and \lx--\lgr~luminosity distributions provide
indications to select candidates for a \g-ray detection,  based on radio and X-ray luminosities. 
The radio luminosity (\lr) alone is not a sufficient selection criterium, in particular for sources with a dominant, luminous (\lr$\gtrsim 10^{42}$ \ergs) jet component.
In fact, for a selected \lr~luminosity in the $\sim 10^{42} - 10^{44}$ \ergs~interval, the corresponding allowed \g-ray luminosities span between $\sim 10^{43}-10^{48}$ \ergs. 
When \lr$\lesssim 10^{42}$ \ergs, the \g-ray luminosities are constrained within an interval of $\sim$1-2 orders of magnitude.
The \lx--\lgr~distribution has a constant linearity over several orders of magnitude.
Nevertheless, radio observations are certainly needed: classification of young radio sources rely on radio observations, which allow us to probe smaller scales than X-rays and to resolve different radio components.\\
\indent
We used the simulated luminosity distributions to derive relations among the three luminosities, \lr~ \lx~and \lgr, for the considered  jet-to-disk ratios and EED. 
A fit of the simulated data applying a linear model, with the coefficients estimated by the least-square method, returned the following relations for the $L_{jet,kin}/L_{disk}=$1.0 distribution in the case of the single power-law EED:
\begin{align}
\Log L_{100MeV-10GeV,44} & =1.34+0.09\times \Log L_{5GHz,44} \nonumber \\
& + 0.94\times \Log L_{2keV,44} ,
\label{eq5}
\end{align}

\noindent
and the broken power-law EED:
\begin{align}
\Log L_{100MeV-10GeV,44} & =0.62+0.48\times \Log L_{5GHz,44} \nonumber \\
& +  0.64\times \Log L_{2keV,44}  ,
\label{eq6}
\end{align}

\noindent
 with all the luminosities in units of $10^{44}$ \ergs.
The scatter in \lgr~at fixed \lr~and \lx~is 0.16 dex and 0.55 dex for Equations \ref{eq5} and \ref{eq6}, respectively. Note that this scatter is derived only from our jet model.\\
\indent
The coefficients of the best fit and relative dispersion for all the simulated distributions are reported in Table \ref{t2}.
We underline that these relations are indicative of the expected \g-ray luminosities for the assumed jet model and assigned ranges of the parameters.

\section{X-ray Young Sources Sample} 

\subsection{Predicted Gamma-ray Fluxes}
In order to search for jet-related \g-ray emission in young radio sources, we selected a sample requiring that: 1- all the sources  have radio and X-ray observations; 2- the jet is likely the most relevant contribution to the non-thermal emission. We considered GPS and CSS radio quasars \citep{Sta01}, as in young radio galaxies the compact radio lobes could dominate the non-thermal contribution \citep{Sta08,Ost10}. 
Our sample is  formed by 13 quasars,  6 classified as GPS sources and 7 as CSS sources. 
All the sources were part of a dedicated X-ray study with \Cha observations. The results of the X-ray analysis were presented in \citet{Sie08}. 
In Table \ref{t3}, we report the redshift, estimated projected size, radio (5 GHz), X-ray (2 keV) and estimated bolometric ($L_{bol}$) luminosities of the sample. The sample, though not complete by any  means, is the largest sample available of GPS and CSS quasars with X-ray (and radio) observations.\\
\indent
Before proceeding further, we comment on the nature of GPS quasars. It is known that GPS quasar samples are partly contaminated by quasars with a core-jet structure which appears compact only because of projection effects. \citet{Sta05} found that large scale emission, indicative of the presence of an extended structure, is often detected in GPS quasars. Even though young quasars could display some degree of radio fluctuations, radio variability gives a definite clue for an aligned quasar, i.e. with a blazar-like nature \citep[see][]{Tin05}. An example is the case of PKS 1127-14, which was  initially classified as a GPS quasar. X-ray and radio observations have revealed the presence of a kiloparsec scale jet \citep{Sie02, Sie07}. This together with the observed multi-band flux variability and superluminal expansion of the core region points to a blazar nature of the source \citep{Bla04}. We still included the source in our sample for comparison with the other GPS/CSS quasars.\\
\indent 
In Figure \ref{f3}, we plot the \lx~of the quasars in the sample as a function of their \lr~and compare them with the simulated
distributions. Here, we have simply ascribed all the 5 GHz luminosity to the jet. This
approximation is more risky in the case of GPS than CSS sources, because
the synchrotron emission of the GPS compact ($\lesssim$1 kpc)
lobes could be important in the GHz-band. 
We have also assumed that the bulk of the 2 keV flux is produced by the jet. This hypothesis is
indeed part of the jet scenario that we want to probe by
comparing model predictions with observations.\\
\indent
The quasars occupy the high-luminosity region of the \lr-\lx~plot. This is well-sampled by our simulations when $L_{jet,kin}/L_{disk}=$1.0 (for both assumed EED), while the overlap with the $L_{jet,kin}/L_{disk}=0.01$ distribution is only marginal. 
A significant overestimate of the \lr~of the observed sample (up to 1-2 orders of magnitude) must be considered to reconcile observations and simulated radio and X-ray luminosities for $L_{jet,kin}/L_{disk}=0.01$. 
On the other hand, the discrepancy with the simulations becomes even larger if only a fraction of the X-ray emission in the quasars of the sample is produced by the jet.\\  
\indent
Interestingly, the simulations show that the lack of a clear correlation between the \lr~and \lx~luminosities of the observed sample does not necessary rule out a common (jet) origin for them.\\
\indent
In Table \ref{t3}, we report the 100 MeV--10 GeV luminosities estimated using Equations \ref{eq5} and \ref{eq6} (we considered only the $L_{jet,kin}/L_{disk}=$1.0 distribution, which better covers the \lr~and \lx~of the sample).   
The predicted \g-ray luminosities span between $\approx$10$^{44}$ and $\approx$10$^{47}$ \ergs, and the minimum and maximum 100 MeV-10 GeV fluxes are $\sim4.2\times10^{-14}$ \ergsc~and $\sim2.5\times10^{-11}$. 
By comparison, the three-year \fermi-LAT flux sensitivity for point sources with 5$\sigma$ detection is $\sim$3$\times$10$^{-12}$ \ergsc~\citep[at Galactic pole and assuming a power-law spectrum with $\Gamma_\gamma=2$, see ][]{Ack12}. 
Seven of the estimated \g-ray fluxes using  Equation (\ref{eq5}) are above or near such threshold, and the number reduces to four for Equation (\ref{eq6}). 
None of the sources in the sample is associated with a \g-ray source in the LAT 2-year Point Source Catalogue \citep[2FGL][ except for PKS 1127-14, associated with the \g-ray source 2FGL1130.3-1448]{Nol12}. However, the 2FGL includes only sources with a minimum $\gtrsim$4 $\sigma$ detection significance\footnote{More specifically, the 2FGL uses the Test Statistics, TS, to quantify how significantly the source emerges from the background and imposes a minimum TS of 25, which corresponds to a significance of just over 4 $\sigma$ evaluated from the $\chi^2$ distribution with 4 degrees of freedom \citep[position and spectral parameters][]{Mat96}.}.

\section{Fermi-LAT Observations}

\subsection{Fermi-LAT Data Analysis}
The {\it Fermi}-LAT is a pair-production \g-ray telescope sensitive to photons in the energy range from 20 MeV to $>$300 GeV. Its large effective area ($\sim$8000 cm$^{-2}$ on axis for $E>$1 GeV) provides view of 2.4 sr of the full sky with an unprecedented angular resolution at these energies (the 68\% containment angles of the reconstructed incoming photon direction below 10 GeV estimated as $\theta_{68}\simeq 0.8^\circ(\varepsilon_{\gamma}/GeV)^{-0.8}$). It mainly operates in a sky-survey mode and covers the full sky every two orbits ($\sim$3 hours). Detailed description of the {\it Fermi}-LAT and of the on-orbit performances are provided in \citet[][]{Atw09}.\\
\indent
We analyzed the {\it Fermi}-LAT data collected during 46 months of operation (August 2008 to February 2012).
For each source, we selected only the events of the Pass 7 V6 {\it
  Source} class, which reduces the residual background rate
\citep[see][]{Ack12}.  We used the Pass7 version 6 of the instrument
response functions \citep[IRFs][]{Ran09,Ack12}.
The data were analyzed with the LAT Science Tools software package (version v9r27p1). We also made use of the LAT Analysis Scripts\footnote{http://fermi.gsfc.nasa.gov/ssc/data/analysis/scitools/\\LATAnalysisScripts.html} (version 0.1.9) for part of the analysis.\\
\indent 
We applied the standard event selections. Time intervals
when the rocking angle of {\it Fermi}-LAT was greater than 52$^\circ$ were
rejected and only events coming from zenith angles smaller than
100$^\circ$ were considered in order to minimize the contribution from
terrestrial albedo
$\gamma$-rays. We considered data in the 0.2 to 100 GeV energy range. Though conservative, this energy selection allow us to reduce systematic errors and the position uncertainties.\\
\indent For each source, we selected a circular region of interest
(RoI) of 10$^\circ$ radius, centered on the radio position of our
candidate.  The full \fermi model that we adopted to calculate an
unbinned likelihood includes all the point-like and diffuse sources
within the RoI. Due to the energy-dependent size of the {\it Fermi}-LAT
point-spread-function (PSF), sources falling between 10$^\circ$ and
15$^\circ$ can contribute to the total counts observed in the RoI and
therefore were also included in the model.  The emission model also
accounts for the Galactic and extragalactic (and instrumental) diffuse
backgrounds. In this work, we used the `mapcube' file
\texttt{gal\_2yearp7v6\_v0.fits} and the \texttt{iso\_p7v6source.txt}
table to describe the emission from the Milky Way and the isotropic
component,
respectively.\\
\indent
We adopted a power-law spectrum ($F=KE^{-\Gamma_{\gamma}}$) for our sources and
assumed the spectral models and parameters reported in the 2FGL
catalog for all the other sources.  
We initially fixed the spectral
parameters of all the sources located 5$^\circ$ away from our
target to the best fit value reported in the 2FGL. We performed a fit
using the unbinned maximum likelihood ({\it gtlike}) and evaluated the
significance of the source detection based on the test statistic
value\footnote{The test statistic is the logarithmic ratio of the
  likelihood of a source being at a given position in a grid to the
  likelihood of the model without the source,
  TS=2log(likelihood$_{src}$/likelihood$_{null}$).}
\citep[TS,][]{Mat96}. 
When the fit did not converge, we simplified the model in the
following way: 1- we fixed the \g-ray photon index of our target to
$\Gamma_{\gamma}=2.5$; 2- we progressively froze the spectral shape
parameters of the sources in the inner 5$^{\circ}$ circle starting
from the farthest ones, as also all the parameters of weak background
sources ($<10^{-14}$ phot cm$^{-2}$ s$^{-1}$ ).

\subsection{Fermi-LAT Results} 
\begin{figure}
\centering
\includegraphics[scale=0.35]{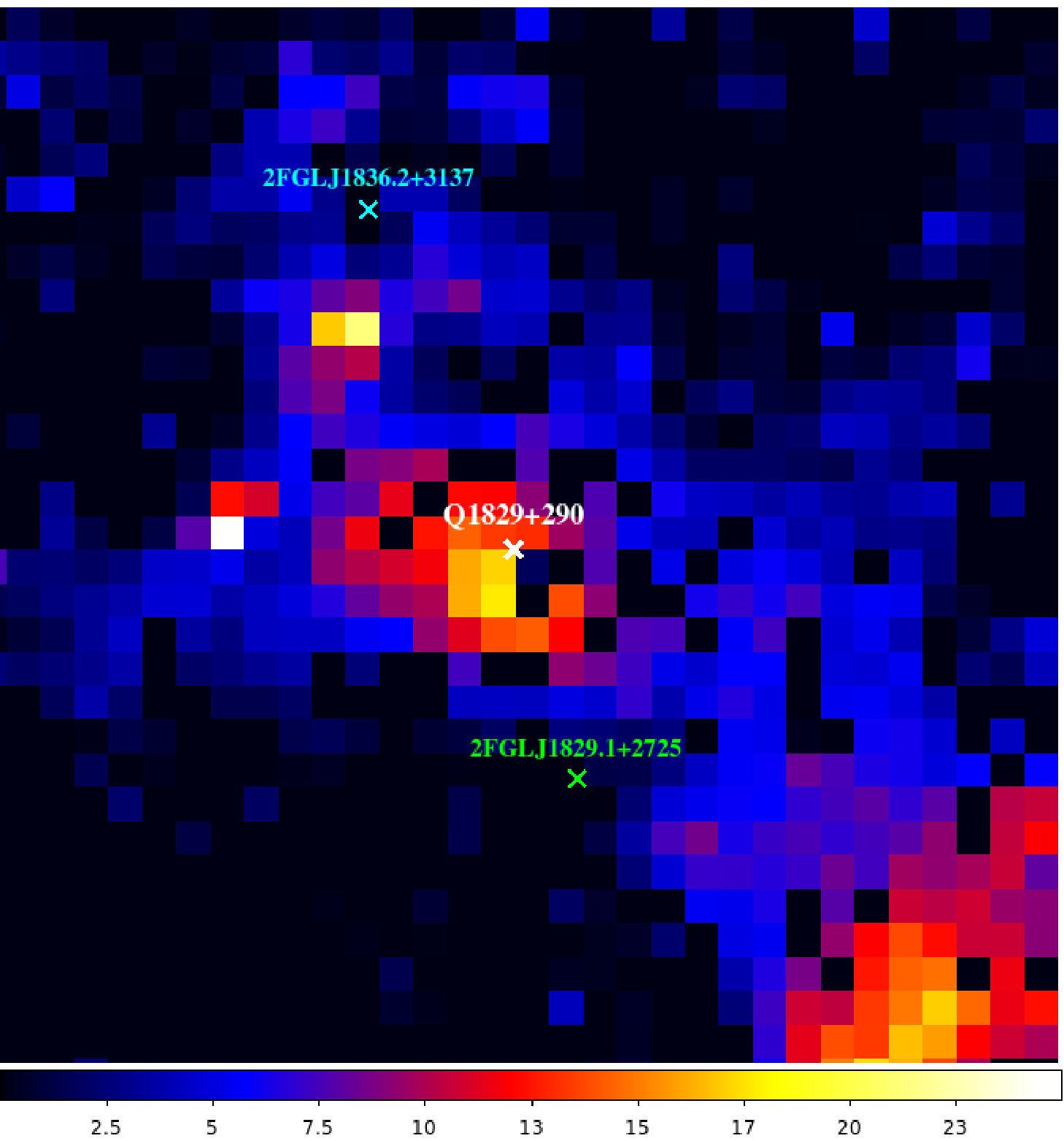}
\includegraphics[scale=0.45]{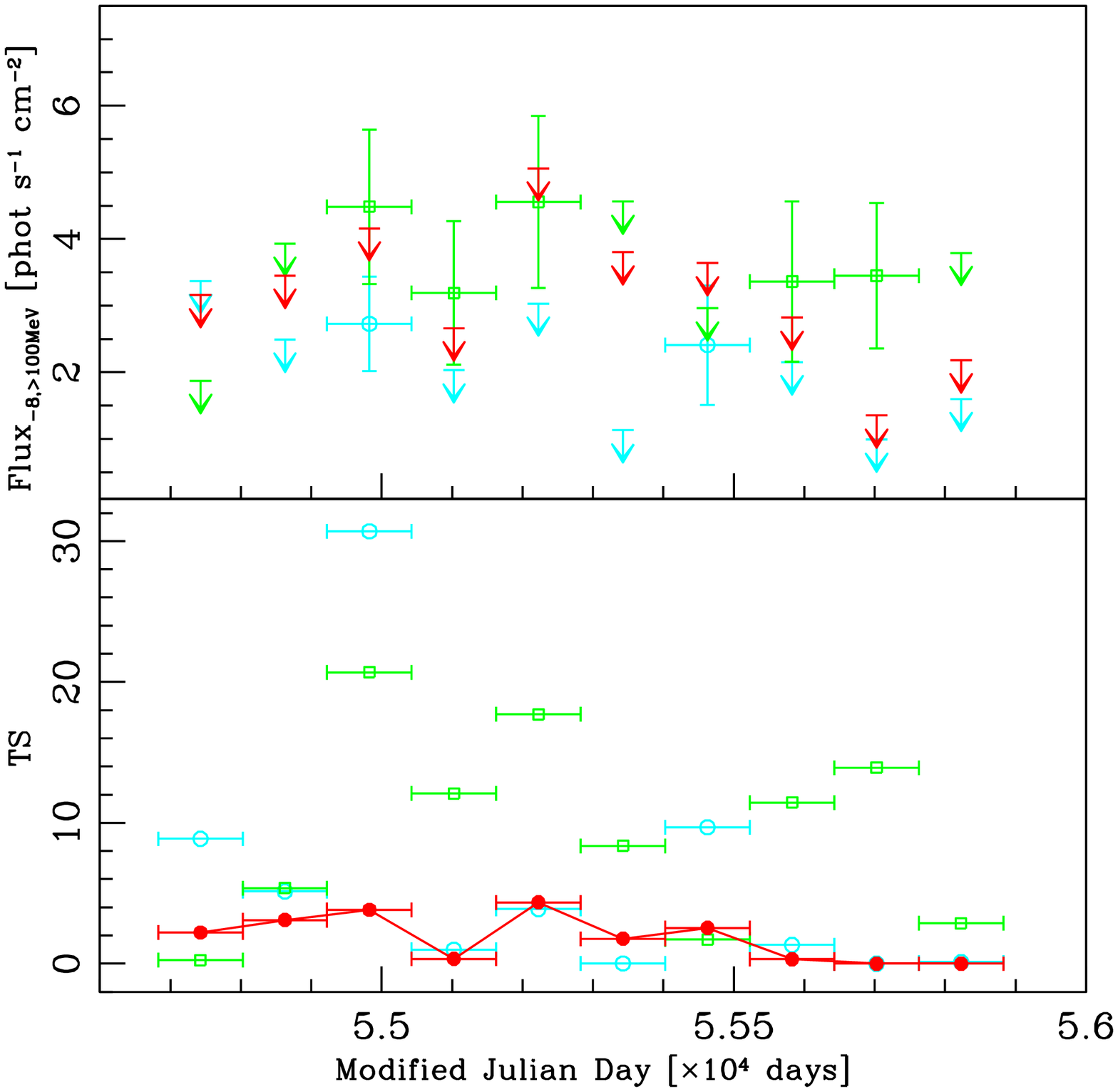}
\caption{Left Panel: {\it Fermi}-LAT TS map of the residuals obtained from the likelihood best fit model without Q1829+290, selecting 32$\times$32 0.25 degree bins. The position of Q1829+290 and the two nearest 2FGL sources are marked in the map. The position of Q1829+290 coincides with a residual excess (TS$\sim$17).
Right Panel: {\it Fermi}-LAT flux lightcurve of Q1829+290 (in red), and of the two nearest $\gamma$-ray sources reported in the 2FGL (2FGLJ1829.1+2725 in green and 2FGL1836.2+3137 in cyan) covering the period from 2008 August 4 to 2012 February 23, using a binning time of 4 months (upper panel). Fluxes are reported in units of 10$^{-8}$ phot cm$^{-2}$ s$^{-1}$ and replaced by upper limit values for TS$<$9. The lower panel shows the corresponding TS values of the 3 sources in the time bins. The TS value of Q1829+290 (in red) is below 9 in all the bins. }
\label{f4}
\end{figure}

The results of the {\it Fermi}-LAT analysis are shown in Table \ref{t4}. In most of the cases, we did not find \g-ray excesses above the background in coincidence with the position of the sources in the sample. For TS values less than 9 ($\lesssim$3 $\sigma$), we calculated the corresponding integral flux upper limit in the range 0.1-10 GeV at 95\% confidence level, setting the photon index of the spectral power-law equal to $\Gamma_{\gamma}=2.5$. \\
\indent
When the \fermi likelihood analysis returned ambiguous results, we further proceeded with the analysis and investigated: 1-
the spatial distribution of the TS value (TS map) in the RoI (using {\it gttsmap}); 2- the temporal distribution of the TS value by producing a 0.2-100 GeV light curve in the selected period probing bin time of months. 
We note that the TS map also allows us to check for the presence of possible new background sources in the 4-year dataset which are not  included in the 2FGL.
In the case of Q1829+290, we obtained a TS value of 20. We built a TS map using the best fit model returned by the likelihood analysis which includes all the sources reported in the 2FGL (without  our source). All the parameters of the model were fixed except for the Galactic diffuse prefactor and the isotropic diffuse normalization. The TS map of the residuals (in Figure \ref{f4}, left panel) shows an excess in the region corresponding to the radio position of Q1829+290 (TS$\sim17$). In the source field there are three 2FGL \g-ray sources within 3\dg~radius (2FGLJ1829.1+2725 at 1.76\dg~distance, 2FGLJ1836.2+3137 at 2.72\dg~and 2FGLJ1842.3+2740 at 2.83\dg ) and, given the {\it Fermi}-LAT PSF in the selected energy band ($\approx$3\dg at 200 MeV), we cannot exclude flux contamination.  We produced the \g-ray lightcurve of Q1829+290 using 4-month time bins and compared it with the lightcurves of the two nearest 2FGL sources. In this case, we left free to vary only the normalization (Integral and Prefactor for the PowerLaw2 and PowerLaw models, respectively)  parameter of the three sources (and again normalizations of the Galactic and isotropic diffuse and instrumental backgrounds). The likelihood analysis was performed in each time interval: the TS values of each of the three sources was computed and for TS$<$9 (3$\sigma$) the flux value was replaced by the upper limit. The flux lightcurves of the three sources, together with their TS values in each time bin, are shown in Figure \ref{f4}. The TS of Q1829+290 is below 9 in all the intervals and its flux fluctuations follow those of the nearest 2FGL source (2FGLJ1829.1+2725), pointing to flux contamination. \\
\indent
PKS1127-14 is the only object associated with a \g-ray source with a high TS value (514). However, the classification of this source has been revised from a GPS to a FSRQ \citep{Hea07}. 
Its \g-ray lightcurve shows variability,  further supporting the interpretation of a blazar-like nature of the core.  We refer to our future paper for a detailed analysis of this source \citep[and see also][]{Aab10a}.  \\
\indent
Summarizing, there was no statistically significant \g-ray detection of the young radio sources in our sample for the considered {\it Fermi}-LAT data set. In the case of Q1829+290, i.e. the only source with a TS value above 9, we cannot exclude flux contamination from near sources in the field.   

\section{Discussion}
The primary goal of the study is to investigate the nature of the high energy emission produced by the jet in young radio sources focusing for the first time on the \g-ray band. The SED simulations have shown that jets of compact sources can be \g-ray luminous: intrinsically powerful jets  observed at moderate angles ($\theta\geq 10$ \dg) could reach \lgr$\sim 10^{48}$  \ergs. The identification of \g-ray luminous young sources is, however, not straightforward, due to our limited knowledge of some of the jet parameters (e.g. $\Gamma$, $\theta$). Radio and X-ray observations are both necessary to select \g-ray candidates among young and bright radio sources (\lr$\gtrsim 10^{42}$ \ergs).  
In the following, we compare the model predictions and observations for the selected sample of young radio quasars and discuss the implications for the jet emission.     
 
\subsection{Model Predictions and LAT Upper Limits}
\begin{figure}
\centering
\includegraphics[scale=0.45, angle=-90]{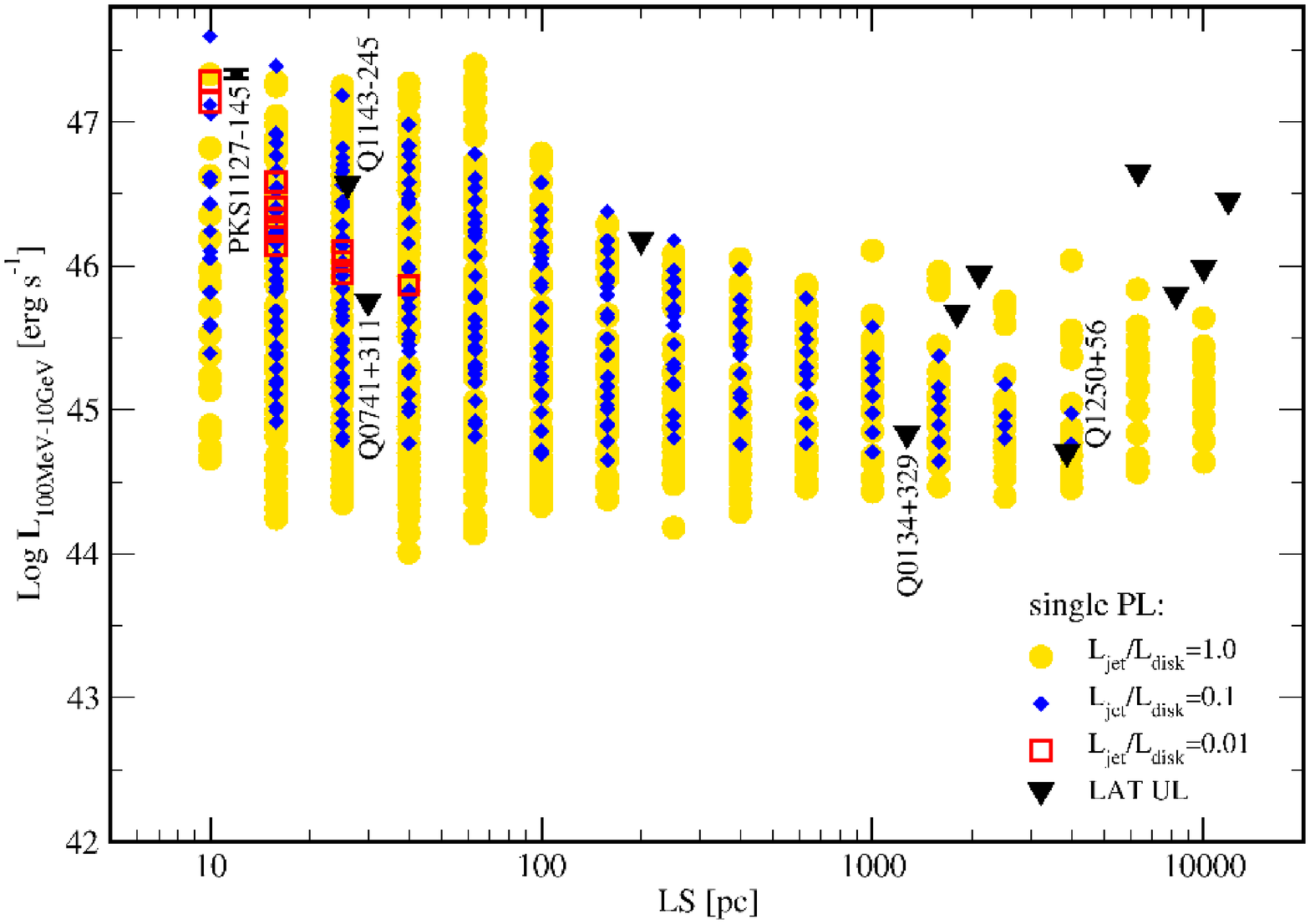}
\vspace{1.0cm}
\includegraphics[scale=0.45, angle=-90]{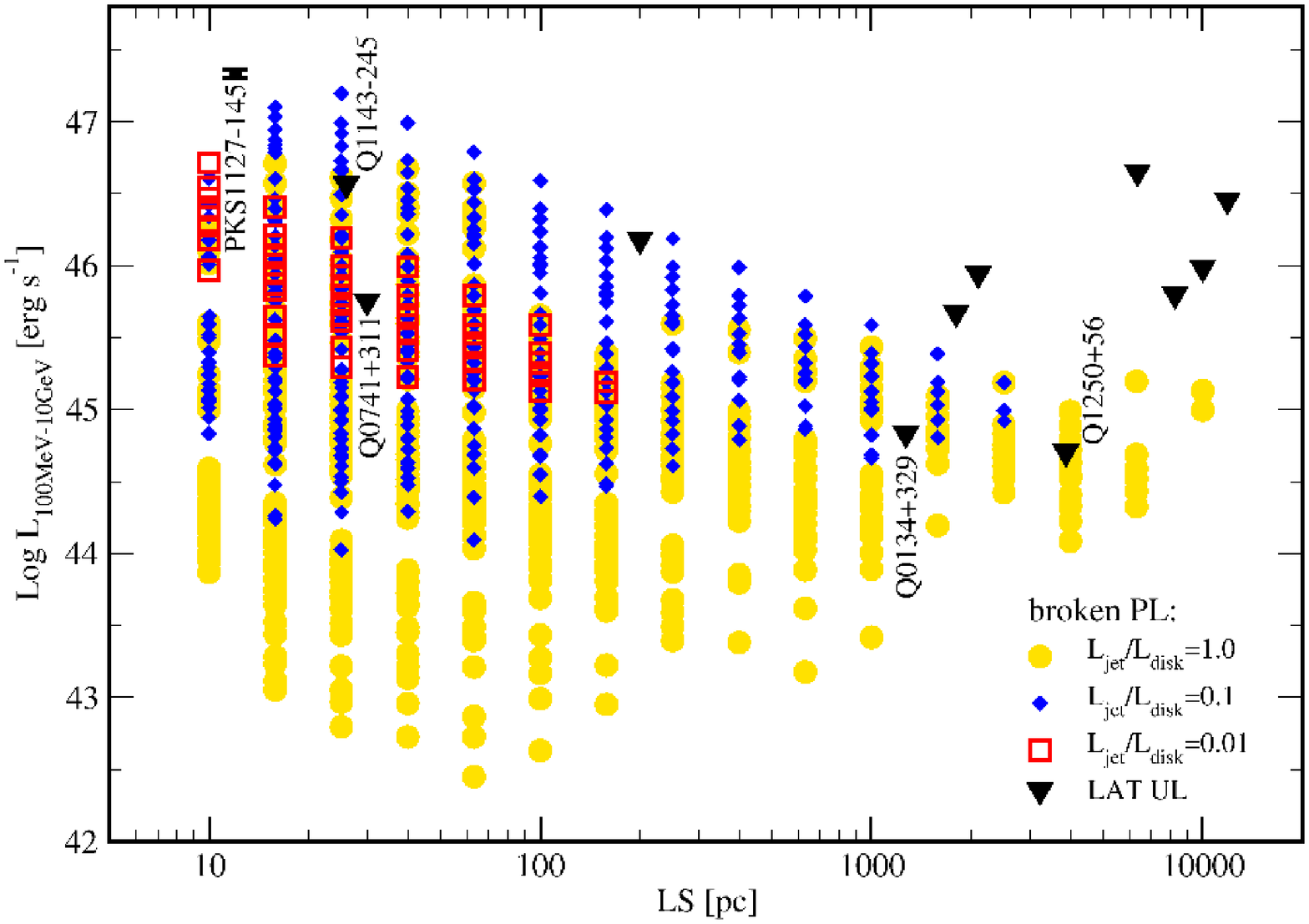}
\caption{Simulated 100 MeV -- 10 GeV luminosities and {\it Fermi}-LAT results are plotted as a function of the linear size, LS. Black solid triangles indicate flux upper limits (calculated between 100 MeV and 10 GeV for a 95\% confidence limit), yellow solid circles the simulated sources assuming $L_{jet,kin}=L_{disk}$, the blue solid diamonds the simulated sources for $L_{jet,kin}=0.1 L_{disk}$ and the red empty squares for $L_{jet,kin}=0.01 L_{disk}$. PKS1127-145 is only detected source by \fermi-LAT. 
Here, a subsample of each simulated distribution is selected  in the range of the observed radio (10$^{43}\lesssim$\lr$\lesssim$10$^{45}$ \ergs) and X-ray (10$^{43}\lesssim$\lx$\lesssim$10$^{46}$ \ergs) luminosities of the GPS/CSS quasar sample (see text). 
Upper panel shows the simulated luminosities assuming a single power law EED and the lower panel assuming a broken power law EED.  }
\label{f5}
\end{figure}

The analysis of the {\it Fermi}-LAT observations did not reveal \g-ray emission associated with the sources in our sample. However, a jet contribution to the high-energy emission is not ruled out. 
In Figure \ref{f5}, the {\it Fermi}-LAT upper limits are compared to the simulated \g-ray luminosities, \lgr, as a function of the linear size, LS (for the observed sample we used the projected jet linear size\footnote{For the rest of the discussion, we excluded from the sample Q1815+6127 which, to our knowledge, has no LS measure reported in the literature.} derived from the radio data and reported in Table \ref{t3}).  
For this comparison, we selected subsamples from each simulated distribution  so that the radio and X-ray luminosities are in the range of the observed sample: 10$^{43}\leq$\lr$\leq$10$^{45}$ \ergs, 10$^{43}\leq$\lx$\leq$10$^{46}$ \ergs~(and 10$^{45}\leq$L$_{bol/disk}\leq$10$^{47}$ \ergs, 10 pc$\leq$LS$\leq$10 kpc).\\
\indent
The luminosity selection determines a progressive reduction of the number of simulated sources with increasing LS. 
Specifically, in the subsamples obtained from $L_{jet,kin}/L_{disk}=$0.1 and 0.01 there are no sources with LS larger than $\sim$5 kpc and $\sim$200 pc, respectively.
The reason is that the radiative efficiency of the jet depends on LS ($L'_{syn}\propto LS^{-1}$, $L'_{EC}\propto LS^{-1}$ and $L'_{SSC}\propto LS^{-2}$), thus for the considered parameter intervals only the most powerful jets ($L_{jet,kin}\gtrsim 10^{46}$ \ergs) can produce the observed radio and X-ray luminosities at kiloparsec scales.\\
\indent
The simulated \g-ray luminosities in the subsamples are compatible with the {\it Fermi}-LAT upper limits, although the {\it Fermi}-LAT sensitivity is the main limit for a test of the model predictions. 
For a few sources the comparison between observations and simulations (in the case of a single power law EED) allows to place constraints on the jet parameters (upper panel in Figure \ref{f5}). 
The {\it Fermi}-LAT limits on Q1143-245 and Q0741+311 rule out the most extreme simulated \g-ray luminosities (\lgr$\gtrsim$10$^{46}-10^{47}$ \ergs). A jet with $L_{jet,kin}/L_{disk}=0.01$ also appears to produce \lgr~above the {\it Fermi}-LAT upper limit of Q0741+311. Similarly, a {\it Fermi}-LAT detection of Q0134+329 and Q1250+56 would be likely if the jet has $L_{jet,kin}/L_{disk}=$1.0 (in the single power law EED case).\\
\indent
The \g-ray luminosity associated with PKS 1127-14, \lgr$=2.16\times10^{47}$ \ergs, is consistent with the highest predicted \g-ray luminosities for the single power-law EED (upper panel in Figure \ref{f5}). The linear size adopted for this source \citep[LS=12 pc, from][]{Ode98} refers to its initial GPS classification (and corresponds to the inner part of the radio jet). 
Such value can be considered as an upper limit for the distance of the emitting region from the black hole. In fact, the location of the \g-ray emitting region in blazar sources is under debate and estimates span from sub- to a few tens of parsecs (see \citet{Sik09} for a discussion and the cases of the quasar PKS 1510-089 and the BL Lacertae object OJ287 in \cite{Mar10} and \cite{Agu11}, respectively).
Interestingly, \citet{Bla04} proposed that the \g-ray emission of PKS 1127-14 is produced via Compton scattering on the IR photons taking place on parsec scales \citep{Bla04}.   
In our simulated distributions, \g-ray luminosities of the order of $10^{46}$-$10^{47}$ \ergs~are obtained via EC/dust in compact (LS$\lesssim$20 pc) and powerful jets ($L_{jet,kin}\gtrsim 10^{46}$ \ergs).  For these parameter values, the emitting region is still moving through the medium which produces the IR photons ($L_{dust}\gtrsim 10^{46}$ \ergs). The highest EC/dust luminosities (in the observer rest frame) are produced for jet bulk motions of $\sim$2-4. For the considered minimum viewing angle  ($\theta$=10\dg), large values of $\Gamma$ (e.g. 10-20) would narrow the radiation cone and reduce the observed luminosity. 
The observed \g-ray luminosity of PKS 1127-14 can be reproduced by assuming a single power law for the electron energy distribution (for all the $L_{jet,kin}/L_{disk}$ ratios), while it is underpredicted in the broken power law scenario (though within an order of magnitude). 
Nonetheless, we note that the high energy SED of blazars inferred from X-ray to \g-ray observations typically displays a complex shape \citep{Aab10a,Gio12}, and a broken power law  appears to be a more realistic description of the energy distribution of the radiating particles.\\ 
\subsection{X-rays: Model Predictions and Observations}
\begin{figure}
\centering
\includegraphics[scale=0.45, angle=-90]{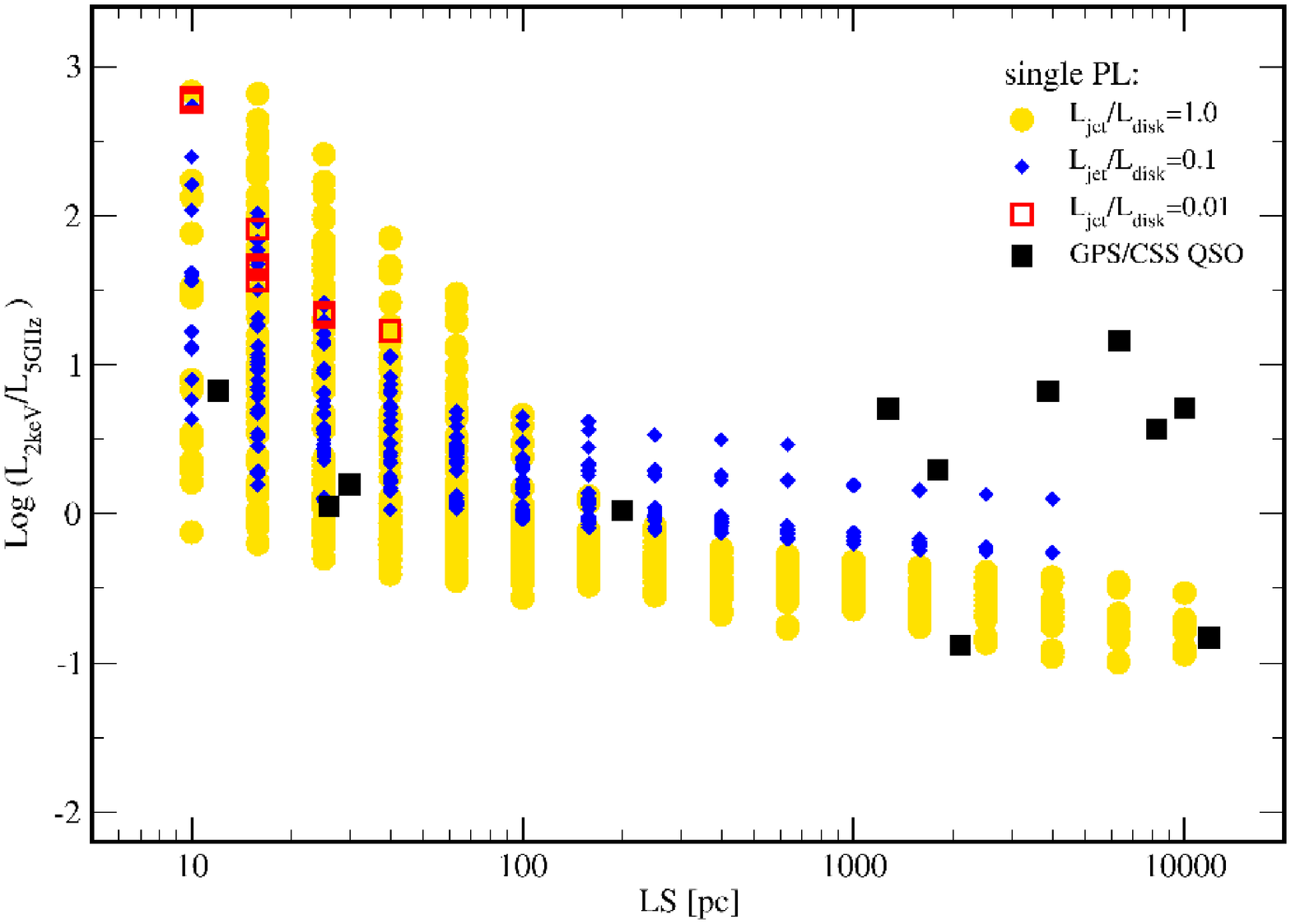}
\vspace{1.0cm}
\includegraphics[scale=0.45, angle=-90]{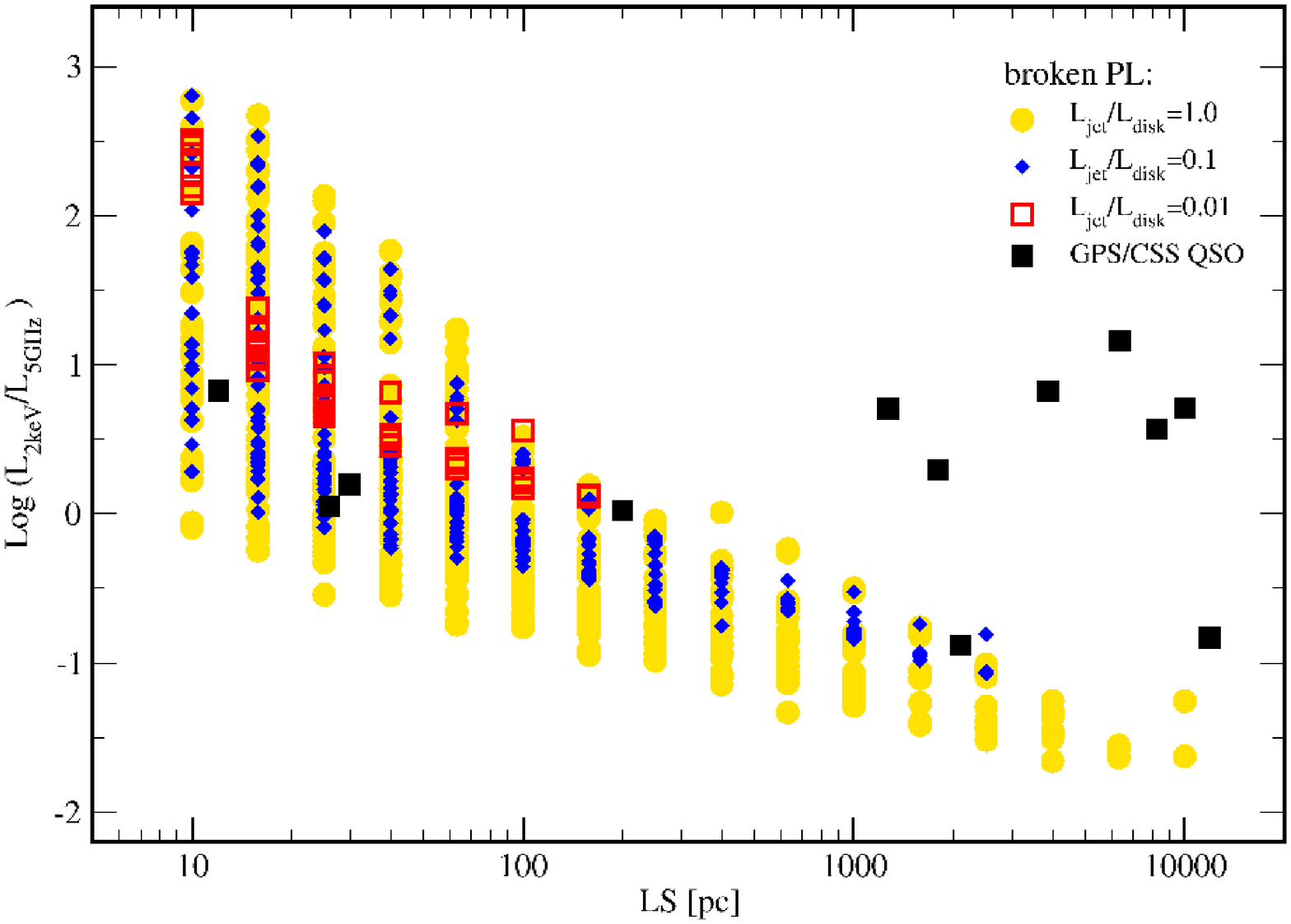}
\caption{ Comparison of the 2 keV to 5 GHz luminosity ratio as a function of the linear size LS for the simulated distributions and the sample of GPS/CSS 
quasars. Yellow solid circles are for simulations with $L_{jet,kin}=L_{disk}$, blue solid diamonds for $L_{jet,kin}/L_{disk}=0.1$ and red empty squares for $L_{jet,kin}/L_{disk}=0.01$. Black solid squares are the GPS/CSS quasars in the sample observed by \Cha \citep{Sie08}. 
Here, a subsample of each simulated distribution is selected in the range of the observed radio (10$^{43}\lesssim$\lr$\lesssim$10$^{45}$ \ergs) and X-ray (10$^{43}\lesssim$\lx$\lesssim$10$^{46}$ \ergs) luminosities of the GPS/CSS quasar sample (see text).
In the upper panel, sources are simulated assuming a single power shape for the EED, in the lower panel a broken power law EED.}
\label{f6}
\end{figure}

We consider the same subsamples of simulated sources to test the model predictions against observations in the X-ray band. In Figure \ref{f6} we show the simulated and observed \lx~to \lr~ratios as a function of LS.
The simulated ratios have a different evolution with LS depending on the shape of the EED.
In the broken power law case the maximum \lx~to \lr~ratio decreases with LS.
Two connected effects cause this trend: 1- at  large LS ($\gtrsim$1 kpc), intrinsically powerful jets ($L_{jet,kin}\gtrsim 10^{46}-10^{47}$ \ergs~and thus $L_{jet,kin}/L_{disk}\geq$0.1) are needed to produce the observed radio and X-ray luminosities of the CSS quasars; 2- the X-ray emission of such powerful jets is dominated by the SSC component, thus \lx/\lr~ scales as $L'_{SSC}/L'_{syn} \propto LS^{-1}$ at all the considered LS.\\
\indent
For the single power law EED, the maximum \lx/\lr~is (roughly) constant for LS$\gtrsim$100 pc (for $L_{jet,kin}/L_{disk}=$1.0, see upper panel in Figure \ref{f6}). Above such LS, the EC/disk and EC/dust contributions to the X-ray band become dominant over the SSC 
(see also Figure \ref{f1}), and thus \lx/\lr$\propto L'_{EC}/L'_{syn} \propto LS^{0}$.
When $L_{jet,kin}/L_{disk}<1.0$, the relative intensity of the disk-related photon fields ($U'_{disk}$ and $U'_{dust}$) with respect to the local synchrotron emission ($U'_{syn}$) increases, therefore the transition to a constant \lx-\lr~ratio  (i.e. $L_{EC}\gtrsim L_{SSC}$ in X-rays) happens at smaller LS than for $L_{jet,kin}/L_{disk}=1.0$.  
\indent
The modeled \lx-\lr~ratios for LS$\lesssim$1 kpc are in good agreement with the ones of the quasar sample. At kiloparsec scales, the model underestimates the \lx--\lr~ ratios observed in CSS quasars (see Figure \ref{f6}). 
The discrepancy is caused by the fact that the simulated sources with LS$\gtrsim 1$ kpc have X-ray luminosities only marginally comparable with, or below, the observed ones but can be radio luminous, with \lr~of the order of $\sim 10^{44}$ \ergs~or larger. 
 This result is not in contradiction with the substantial overlap between simulated and observed sources shown in Figures \ref{f3}. In fact, in that case the comparison is limited to two observables only (\lr~and\lx), while in Figure \ref{f6} we introduce additional information (i.e the linear size and the relation between the radio and X-ray jet luminosities, i.e. \lx/\lr).\\
The  gap between the observed and simulated sources, though still present, is reduced for the subsamples  simulated assuming a single power law EED and when $L_{jet}/L_{disk}=0.1$. However, in the $L_{jet,kin}/L_{disk}=0.1$ distribution the number of kiloparsec scale sources, which abide by the subsample selection criteria, is limited.
We also note that the difference between simulated and observed \lx--\lr~ratios could be even more severe when we consider the true, i.e. not the projected, linear size. \\
 
\subsubsection{Parameter values \& Assumptions} 
\begin{figure}
\centering
\includegraphics[scale=0.6, angle=-90]{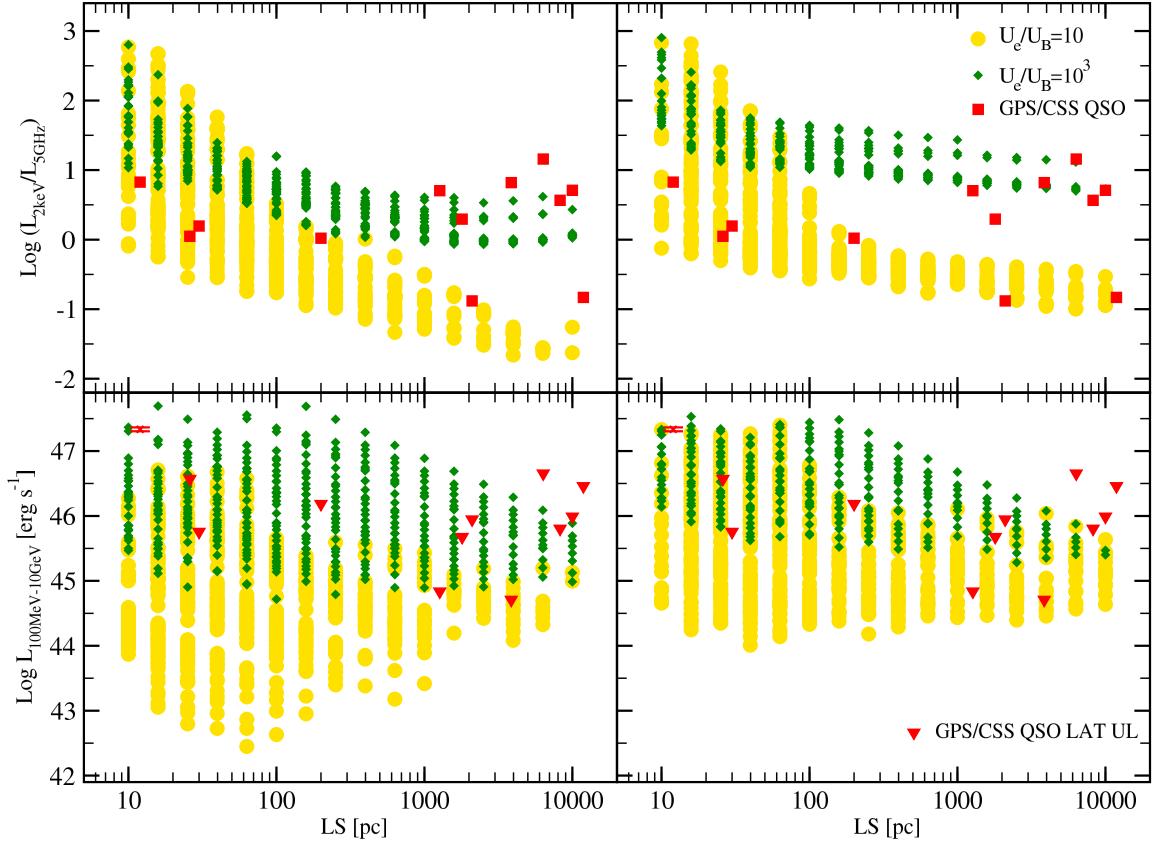}
\caption{Upper Panels: \lx~to \lr~ratios are shown as a function of the linear size LS for simulated sources and the sample of GPS/CSS quasars (red solid squares). Lower Panels: Simulated 100 MeV -- 10 GeV luminosities and \fermi-LAT upper limits (red solid triangles) plotted as a function of the linear size, LS. Left panels show simulations assuming a broken power law EED and right panel a single power law EED. Yellow solid circles are the simulated sources when $U_e/U_B=10$ while the green solid diamonds for $U_e/U_B=10^3$. In all simulations $L_{jet,kin}=L_{disk}$ is assumed. Subsamples have been selected from the distributions as described in the text.  }
\label{f7}
\end{figure}

Bearing in mind the limited size of our sample, we can discuss the differences between the model and observations. We first consider whether different parameter values and assumptions might change the predictions. \\
A more compact emitting region than the one assumed here ($R=0.1 LS$) would increase the SSC luminosity. 
We tested the case of R=0.01$\times$LS. The \lx-\lr~ratios increase but at kiloparsec scales they are still below the observed ratios.\\ 
\indent
Can different Doppler factors, $\delta$, modify the model results in the right directions? In our simulations $\delta$ is between $\sim$0.3 and $\sim$5. For the considered values of $\theta$ (10\dg-50\dg), which reasonably sample the viewing angles for quasars, $\Gamma$ above the maximum assigned value of 10 would not increase $\delta$. Furthermore, the observed SSC and synchrotron flux densities have the same dependency on $\delta$. Thus this factor cannot significantly modify the \lr/\lx~of jets where the SSC luminosity dominates the X-ray emission. The boosting of the external Compton luminosity depends on the arrival direction of the seed photons in the jet comoving frame \citep{Der95}. If the emitting region is still within the putative torus region, for instance, there is a $\sim\delta^{1+\alpha}$ factor between the EC/dust and the synchrotron observed luminosities. However, for the considered external fields, such effect could be important for small size sources (tens of parsec), without affecting the ratios in the CSS sample. \\
\indent
The discrepancy between observations and simulations holds true for all the simulated subsamples, though to different amplitude, suggesting that it does not strongly depend on the assumed spectral shape of the electron distribution.\\
\indent
An increase of the particle to magnetic field energy density ratio is a viable option. In Figure \ref{f7}, we show that simulations assuming $U'_e/U'_B\gtrsim$10$^3$ (and $L_{jet,kin}=L_{disk}$) can reproduce the observed \lx--\lr~ ratios. Interestingly, the corresponding simulated $\gamma$-ray luminosities for the majority of the sources are above the measured {\it Fermi}-LAT upper limits and would make a \g-ray detection likely.\\

\subsection{Origin of the X-ray emission} 
The study of the parameters seems to confirm that the observed X-ray emission can be hardly produced by the jet at scales larger than $\sim$1 kpc.
This can be interpreted in two ways: 1- the bulk of the jet X-ray emission is always produced at distances smaller than $\sim$1 kpc from the central BH; 2- the observed X-ray emission is due to an additional component not related to the jet, such as the disk-corona emission. \\
\indent 
In our model, the dissipation region moves along the jet as the source expands: the decrease of the X-ray emission with LS is a consequence of the increasing volume of the emitting region and decrease of the intensities of the nuclear photon field ($\propto1/LS^2$). 
This is certainly a simplified picture, since large ($>$10 kpc) radio jets may display multiple X-ray knots and diffuse X-ray emission.  
 However, in our simulations the case of a bright knot located at an intermediate position along an extended jet is accounted for by considering a distribution of linear sizes\footnote{We do not assume any condition specific to the jet termination for the knot emission.} (e.g. a knot located at 100 pc in a 10 kpc jet will have the same luminosity of a 100 pc jet with the same jet power). Continuous X-ray emission is typically present at lower luminosity levels, and here we aim at modeling the brightest X-ray component of a jet.\\
\indent
\citet{Sta05} found that in GPS quasars the origin of the radio emission appears to be close to the core.  
This means that either the bulk of the emission is produced by the inner part of the jet or, as pointed out by the authors, it is due to projection effects making the compact and young nature of these sources uncertain.  
\citet{Sal09} investigated the origin of the X-ray emission observed in the GPS quasar 3C~287 and concluded that it is likely associated with the inner radio jet, whose axis is closely aligned to the observer line of sight. The authors argue that the lack of variability, which is typical of blazar sources, might be related to particular physical conditions of the jet in the initial phase. In this scenario, the non detection with {\it Fermi}-LAT can also be in favor of a non-blazar nature. 
In the CSS quasar 3C~48, \citet{Wor04} reported the presence of a hard X-ray excess which is ascribed to the inner part of the jet and a similar explanation is invoked for the CSS source PKS 2004-447 \citep{Gal06}. \\
\indent 
As an alternative, the X-ray emission may be primarily contributed by thermal radiation from the innermost region of the accretion disk or Comptonization of the disk photons in a hot corona as observed in the X-ray cores of quasars and FRII radio galaxies. 
\citet{Sie08} found that the median optical-to-X-ray luminosity parameter ($\alpha_{oX}$) for their sample of GPS and CSS quasars and radio galaxies  is 1.53$\pm$0.24, in agreement with $\alpha_{oX}$ of radio-quiet quasars \citep{Kel07a}. In addition, the estimated disk luminosities of the observed quasars suggest a radiatively efficient disk. An accretion-related origin is also plausible for the X-ray emission of GPS radio galaxies, which appears to be highly absorbed \citep[with equivalent hydrogen column density of $N_H>10^{22}$ cm$^{-2}$,][]{Gua06,Vin06,Sie08,Ten09}. An alternative explanation in the framework of non-thermal emission from the lobes is proposed by \citet{Sta08} and \citet{Ost10}.\\
\subsection{Comparison with Gamma-ray detected Misaligned AGN}
A comparison with non-blazar radio sources detected in \g-rays is useful to understand how the jet emission can contribute the X-ray to \g-ray band. Here, we consider the recent \g-ray studies on misaligned AGN (MAGN) and Broad Line Radio Galaxies \citep[BLRG,][]{Aab10,Kat11,Gra12}. The sample of  MAGN observed by {\it Fermi}-LAT is dominated by low-power FRI radio sources, while powerful FRII appear to be an elusive \g-ray class.  
The different redshift distribution of the two classes, with FRII being (typically) more distant, seems not sufficient to explain the low rate of detection. An intrinsic difference in the jet structure and/or emission mechanism responsible for the \g-rays has been proposed as a possible explanation \citep[][and references therein]{Gra12a}. Jets with a complex dynamical structure as well as beaming difference between SSC and EC \citep{GK03,GTC05,Gia10} processes can account for a smaller number of FRII sources than FRI. Among the 18 BLRG investigated in \citet{Kat11} using two-year \fermi-LAT data, only 3C 111 and 3C 120 have \g-ray detections. In these sources, the \g-ray emission is ascribed to the unresolved cores and  displays variability on month (or longer) timescales, with the consequence that the sources are not always \g-ray detected.\\
\indent
The sources in our sample are all radio powerful and most of them morphologically resemble FRII radio sources. Thus, it is possible that they share the same \g-ray elusiveness. The results of our study are in agreement with previous findings that the jet does not likely dominate the X-ray core emission. In BLRG cores, the  jet emission can contribute to some fraction of the X-ray emission but it is typically hidden by the emission of the accretion disk and corona  \citep{Zdz01,Gra07,Kat07,Sam09}. In the 2-10 keV band, \citet{Gra07} constrain the jet to accretion flow relative contribution to a factor of $<$0.7.  Nonetheless, the jet might emerge at higher energies, as confirmed by the \g-ray detections of 3C 111 and 3C 120. \citet{Kat11} evaluate the jet and disk relative contributions in the selected BLRG using a composite model of a thermal  plus non-thermal templates to reproduce the broadband core emission \citep[similar to the approach adopted  in][]{Gra07}. The authors conclude that the total (observed) luminosity of the jet is on average not less than 1\% of the disk-corona luminosity, and in some sources can even be almost comparable to it. 
 
\section{Summary}
\begin{figure}
\centering
\includegraphics[scale=0.4, angle=-90]{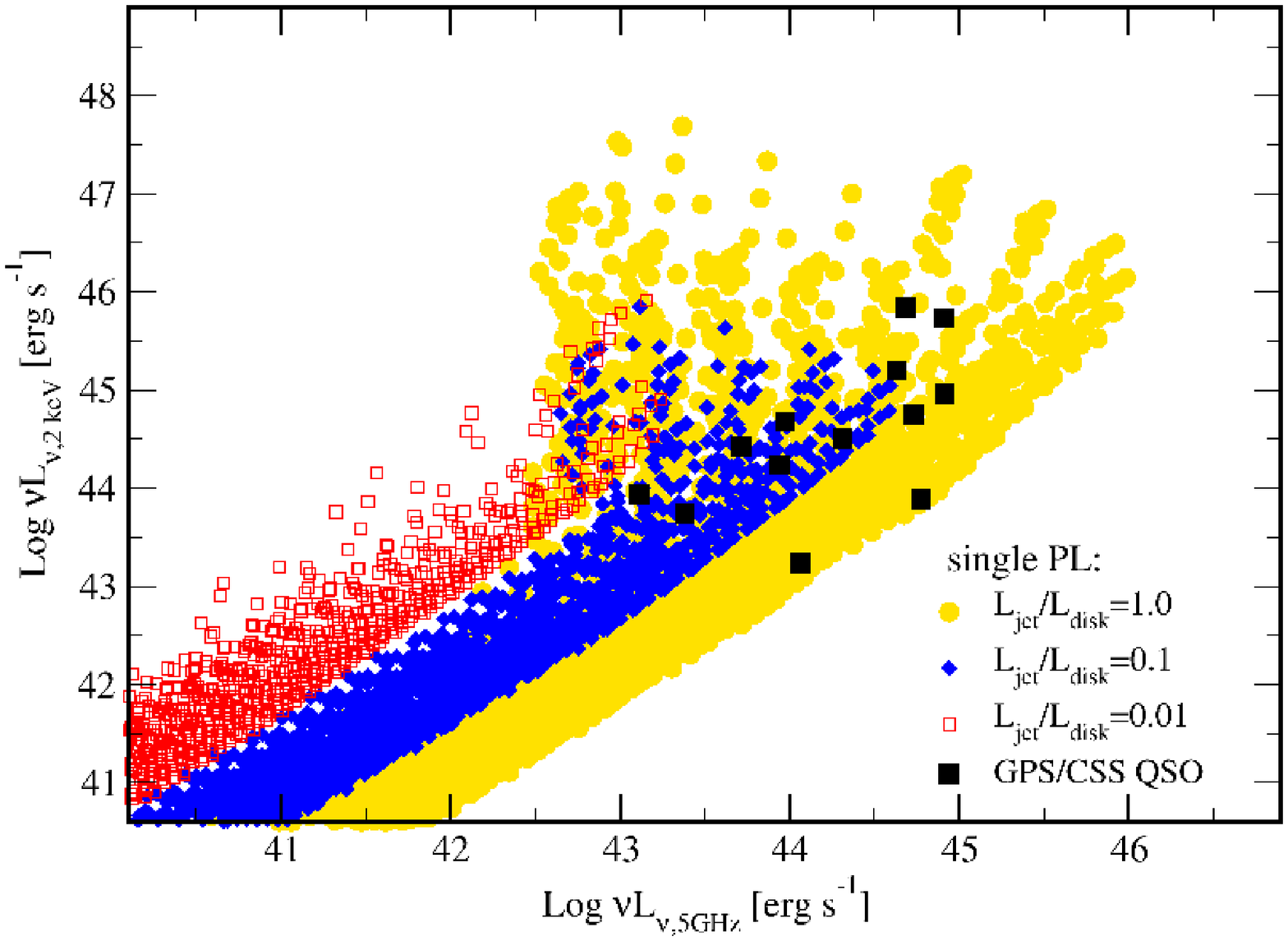}
\hspace{0.2cm}
\includegraphics[scale=0.4, angle=-90]{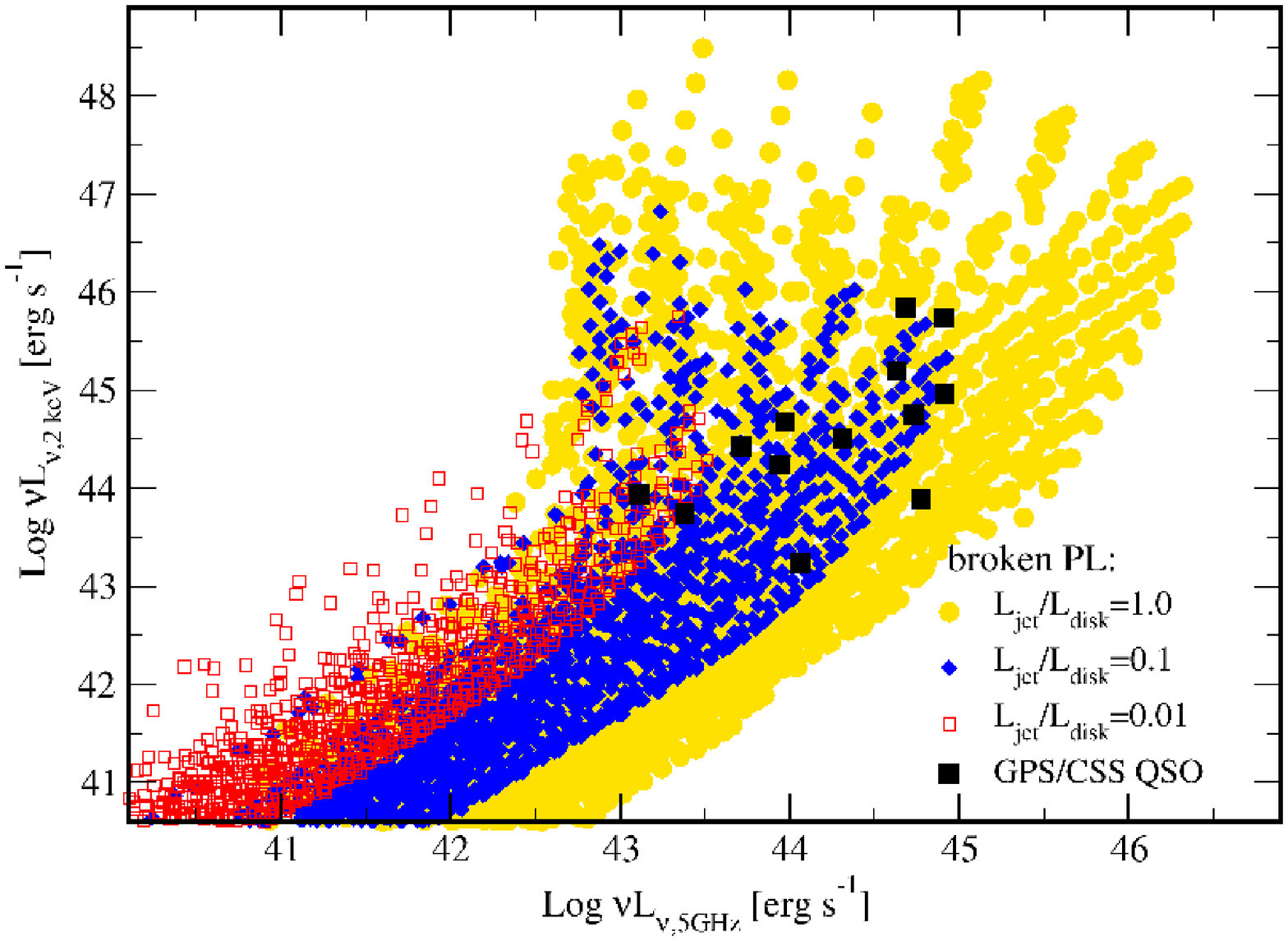}
\caption{X-ray luminosities \lx~of the GPS and CSS quasars (Table \ref{t3}) and the simulated sources as a function of radio luminosities \lr~    (observer rest frame): yellow solid circles are the simulated sources assuming $L_{jet,kin}=L_{disk}$, the blue solid diamonds the simulated sources for $L_{jet,kin}=0.1 L_{disk}$ and the red empty squares for $L_{jet,kin}=0.01 L_{disk}$. Black solid squares are the GPS and CSS quasars.}
\label{f3}
\end{figure}

We investigated the high energy emission produced by jets in young radio quasars, focusing on a sample of GPS and CSS quasars previously detected in X-rays by \Cha. The analysis of $\sim$46 month {\it Fermi}-LAT data does not reveal any statistically significant \g-ray detection of the sources, except for the already known case of PKS1127-14. 
For the first time, we compared the {\it Fermi}-LAT and \Cha observations of the sample to \g-ray and X-ray luminosities predicted by a simple jet synchrotron and IC radiative model.
The simulations performed for a reasonable set of model parameters and assumptions have shown that for the analyzed sample:
\begin{itemize}
\item[(i)]{$L_{jet,kin}/L_{disk}>$0.01 is required for the range of disk luminosities of the sources: low power jets (as for the $L_{jet,kin}/L_{disk}=$0.01 case) fail to produce the majority of the observed radio luminosities (Figure \ref{f3});}
\item[(ii)]{a jet origin of the bulk of the observed X-ray emission is strongly disfavored. Specifically, this holds true for X-ray emission  associated with kiloparsec jets (e.g. jets of CSS quasars) even when we significantly change our initial assumptions; }
\item[(iii)]{large deviations from energy equipartition ($U'_e/U'_B>>10$), necessary to produce the observed X-ray luminosities, would imply \g-ray fluxes detectable by \fermi-LAT.}
\end{itemize}
Our analysis indicates that in the considered sample the bulk of the X-ray emission is either related to the inner ($<$1 kpc) jet or to a different component, most likely the disk-corona.\\
\indent 
This study also shows that the \g-ray band can be important for the classification of compact radio sources, to distinguish between young and aligned sources: a GPS quasar that is radio and X-ray bright but not detected in \g-rays (at the current level of the LAT al sky-survey sensitivity) is likely to be truly a young compact radio source rather than a blazar with large radio structures observed at small viewing angles.

\begin{acknowledgments}
The authors thank M. Sikora and D.~E. Harris for useful suggestions and comments. We are grateful to the anonymous referee for critical comments which helped to improve the paper.
This work was supported by the National Aeronautics and Space Administration contract NAS8-03060 and under the following grants: NNX10AO60G issued through the Fermi Guest Observer Program and GO1-12145X issued through the Chandra X-ray Observatory Guest Observer program. 	
The authors thank the Fermi Science Support Center Help Desk team for support in the data analysis. 
This research has made used of TOPCAT\footnote{http://www.star.bris.ac.uk/$\sim$mbt/topcat/} \citep{Tay05} for the preparation and manipulation of the tabular data.
\end{acknowledgments}

\newpage

\begin{table}
\caption{Model Parameters}
\label{t1}
\begin{center}
\medskip
\begin{tabular}{l l c}
\hline
\hline
\multicolumn{1}{l}{Notation}&
\multicolumn{1}{l}{Definition}&
\multicolumn{1}{c}{Value/Range}\\
\\
\hline
\\
\multicolumn{3}{c}{Free Input Parameters}
\\
LS                 & jet linear size (core to jet termination)        & 10 pc--10 kpc\\
$L_{disk}$    &bolometric disk luminosity    &10$^{45}$--10$^{47}$ \ergs\\
$\Gamma$   &jet knot bulk Lorentz factor   &1.4--10\\
$\theta$          &observer's viewing angle       &10\dg--50\dg\\
\hline
\\
\multicolumn{3}{c}{Linked Parmeters}
\\
R			    & knot radius            & 0.1LS\\
z$_d$            & knot distance from the BH  &LS\\
$L_{jet,kin}$    &jet kinetic power           & (0.01,0.1,1.0)$L_{disk}$\\
$U'_e$        &radiating electron energy density      &$\propto L_{jet,kin}$\\
$U'_B$               &magnetic field energy density        &$U_e/U_B=10$\\
$L_{dust}$    &dust/IR luminosity                 &0.1$L_{disk}$\\
$R_{dust}$   &dust shell radius                      &$\propto\sqrt{L_{disk}}$\\
$L_{bl}$        &synchrotron radiative power of the blazar-like blob &$10\%L_{jet,kin}$ \\
\hline
\\
\multicolumn{3}{c}{Fixed Parmeters}
\\
\g$_{min}$ &minimum electron Lorentz factor &10\\
\g$_{max}$ &maximum electron Lorentz factor &10$^5$\\
\g$_{break}$ &energy break Lorentz factor (broken power law) &$2\times10^3$\\
$p$               &EED index (single power law)               &2.68\\
                     &EED low energy index (broken power law) &2.1\\
$p2$             &EED high energy index (broken power law) &4.0\\
T$_{disk}$    &disk blackbody temperature  &3$\times$10$^4$ K\\
T$_{dust}$    &dust blackbody temperature  &370 K \\
$\Gamma_{bl}$ &blazar-like blob bulk Lorentz factor &10\\
\hline\end{tabular}
\end{center}
\end{table}

\begin{table}
\caption{Results of the least square model fit for the \lr-\lx-\lgr~ simulations.}
\label{t2}
\begin{center}
\medskip
\begin{tabular}{l l | c c c c}
\hline
\hline
  \multicolumn{2}{c|}{Simulations} &
  \multicolumn{4}{c}{Best Fit}\\
  \multicolumn{1}{l}{$L_{jet,kin}/L_{disk}$} &
  \multicolumn{1}{l|}{EED} &
  \multicolumn{1}{c}{x0} &
  \multicolumn{1}{c}{x1} &
  \multicolumn{1}{c}{x2} &
  \multicolumn{1}{c}{disp.}\\
   \multicolumn{1}{l}{(1)}&
  \multicolumn{1}{l|}{(2)}&
  \multicolumn{1}{c}{(3)}&
  \multicolumn{1}{c}{(4)}&
  \multicolumn{1}{c}{(5)}&
  \multicolumn{1}{c}{(6)}\\
&&&&&\\
\hline
&&&&&\\
1.0         & SP     &1.34    &0.09   &0.94   &0.16\\
            & BP     &0.62    &0.48   &0.64   &0.55\\
&&&&&\\
0.1         & SP     &1.63    &0.05   &0.99   &0.10\\ 
            & BP     &1.51    &0.41   &0.72   &0.40\\
&&&&&\\
0.01        & SP     &1.68    &0.07   &0.97   &0.12\\
            & BP     &1.02    &0.21   &0.91   &0.40\\
\hline\end{tabular}
\end{center}
Columns:1- assumed jet-to-disk ratio; 2- shape of the EED: SP=single power law, BP=broken power law; 3- intercept (\lgr~are in units of $10^{44}$ \ergs); 4- $L_{5GHz,44}$ coefficient; 5- $L_{2keV,44}$ coefficient; 6- intrinsic dispersion.   
\end{table}

\begin{table}
\caption{The sample of GPS/CSS quasars with \Cha observations in \citet{Sie08}.}
\label{t3}
\begin{center}
\medskip
\begin{tabular}{l r r r r c c c}
\hline
\hline
  \multicolumn{1}{l}{Source} &
  \multicolumn{1}{c}{Type}&
  \multicolumn{1}{c}{z} &
  \multicolumn{1}{c}{LS}&
  \multicolumn{1}{c}{log(L$_{5GHz}$)} &
  \multicolumn{1}{c}{log(L$_{2keV})$} &
  \multicolumn{1}{c}{log(L$_{bol})$} &
  \multicolumn{1}{c}{log(L$_{100-10,SPL}$)$^a$/log(L$_{100-10,BPL}$)$^b$} \\
  \multicolumn{1}{c}{} &
  \multicolumn{1}{c}{} &
  \multicolumn{1}{c}{} &
  \multicolumn{1}{c}{pc} &
  \multicolumn{1}{c}{erg/s} &
  \multicolumn{1}{c}{erg/s} &
  \multicolumn{1}{c}{erg/s} &
  \multicolumn{1}{c}{erg/s} \\
  \multicolumn{1}{l}{(1)}&
  \multicolumn{1}{c}{(2)}&
  \multicolumn{1}{c}{(3)}&
  \multicolumn{1}{c}{(4)}&
  \multicolumn{1}{c}{(5)}&
  \multicolumn{1}{c}{(6)}&
  \multicolumn{1}{c}{(7)}&
  \multicolumn{1}{c}{(8)}\\
\\
\hline
\\
  Q0134+329/3C~48      & CSS              & 0.367         & 1.27$\times$10$^3$       & 43.97        & 44.68   &45.96     & 45.98/45.04 \\
  Q0615+820                  & GPS              & 0.71           & 1.80$\times$10$^3$       & 43.94        & 44.23    &45.37     & 45.56/44.74 \\
  Q0740+380/3C~186    & CSS              & 1.063         & 10$\times$10$^3$          & 43.72        & 44.42   &46.02     & 45.71/44.75  \\
  Q0741+311                  & GPS              & 0.63           & 30                                   & 44.31        & 44.51        &45.11     &45.85/45.09 \\ 
  Q1127-145                   & GPS$^\ast$  & 1.18           & 12                                   & 44.91        & 45.74       &46.80     & 47.06/46.17 \\
  Q1143-245                   & GPS              & 1.95           & 26                                   & 44.91        & 44.96        &45.75     & 46.32/45.67 \\
  Q1245-197                   & GPS              & 1.28           & 2.1$\times$10$^3$         & 44.77        & 43.89     &44.80     & 45.31/44.91 \\
  Q1250+56/3C~277.1   & CSS              & 0.32           & 3.86$\times$10$^3$       & 43.11        & 43.93   &45.23     & 45.19/44.15\\
  Q1328+254/3C~287    & CSS              & 1.055         & 200                                 & 44.73         & 44.75     &46.10     & 46.11/45.45 \\
  Q1416+067/3C~298    & CSS              & 1.439         & 6.35$\times$10$^3$       & 44.68         & 45.84  &46.90     & 47.13/46.12 \\
  Q1458+718/3C~309.1 & CSS              & 0.905         & 8.25$\times$10$^3$       & 44.63         & 45.20  &47.14     & 46.52/45.69 \\
  Q1815+614                  & GPS              & 0.601         & --                                     & 43.38         & 43.74       & 44.70       & 45.04/44.16 \\
  Q1829+290                  & CSS              & 0.842         & 11.86$\times$10$^3$     & 44.06         & 43.23    &44.96    & 44.63/44.16 \\
\hline\end{tabular}
\end{center}
Columns: 1 -- Source name; 2 --  GPS or CSS radio classification based on  \citet{Ode98}; 3 -- redshift; 4 -- projected jet linear size (calculated from \citet{Ode98} with current cosmology); 5 -- radio luminosity at 5 GHz \citep[see][and references therein]{Sie08}; 6 -- X-ray luminosity at 2 keV \citep{Sie08}; 7 -- bolometric luminosity; 8 -- predicted $\gamma$-ray (100 MeV--10 GeV) luminosity (see text).\\
$^a$: calculated using Equation \ref{eq5} for a single power law EED, the scatter is 0.16 (dex). $^b$: calculated using Equation \ref{eq6} for a broken power law EED, the scatter is 0.55 (dex). $^\ast$: classification for this source has been revised (see text).
\end{table}

\begin{table}
\caption{ GPS/CSS sample -- {\it Fermi}--LAT Results}
\label{t4}
\begin{center}
\medskip
\begin{tabular}{lccc}
\hline
\hline

Source                &TS$^a$   & {\it Fermi}-LAT Flux$^b$     &{\it Fermi}-LAT Flux         \\
                           &                &$\times10^{-9}$ phot cm$^{-2}$ s$^{-1}$     &$\times10^{-12}$ erg cm$^{-2}$ s$^{-1}$      \\
(1)    &(2)       &(3)   &(4)\\                           
\\
\hline
\\
Q0134+329                &$<$0       &$<$3.5     &$<1.51$            \\

Q0615+820                &3         &$<$4.9      &$<2.12$         \\

Q0740+380                &3         &$<$3.7      &$<1.60$        \\
 
Q0741+311                &6         &$<$7.9    &$<3.41$              \\

PKS1127-145            &514        &63$\pm$4$^c$   &27.30\er1.73    \\

Q1143-245                 &1         &$<$3.1     &$<1.34$          \\

Q1245-197                 &$<$0        &$<$2.1    &$<0.91$      \\

Q1250+568                &1        &$<$3.6     &$<1.56$          \\

Q1328+254                &1       &$<$5.9        &$<2.55$  \\

Q1416+067                &3        &$<$8.0       &$<3.46$   \\

Q1458+718                &1        &$<$3.6       &$<1.56$   \\

Q1815+6127              &$<$0       &$<$1.2      &$<0.52$    \\

Q1829+290                &20         &$<$19.5$^d$   &$<8.42$       \\

\hline
\end{tabular}
\end{center}
Columns: 1 -- Source name; 2 -- Test statistic value; 3,4 -- {\it Fermi}--LAT 100 MeV--10 GeV fluxes.\\ 
$^a$: when TS$<$9 ($\approx 3\sigma$), the reported flux is an upper limit. 
$^b$: upper limits are calculated in the 100 MeV - 10 GeV energy band, for a fixed \g-ray spectral index of 2.5, at 95\% confidence interval.
$^c$: the measured \g-ray spectral index is 2.75\er0.05. 
$^d$: contamination of nearby ($<$3 \dg) \g-ray sources cannot be excluded (see text and Figure \ref{f4}), thus even in this case we report to upper limit value.

\end{table}

\end{document}